\documentclass[sensors,article,accept,pdftex,moreauthors]{Definitions/mdpi} 
\firstpage{1} 
\pubvolume{1}
\issuenum{1}
\articlenumber{0}
\pubyear{2024}
\copyrightyear{2024}
\externaleditor{Academic Editor: Firstname\linebreak Lastname}
\datereceived{23 May 2024} 
\daterevised{6 June 2024} 
\dateaccepted{ } 
\datepublished{ } 
\hreflink{https://doi.org/} 

\usepackage[labelformat=simple]{subcaption}

\DeclareCaptionLabelFormat{subcaptionlabel}{\normalfont(\textbf{#2}\normalfont)}
\usepackage{amsmath}
\usepackage{amssymb}
\usepackage{amsfonts}
\usepackage{array}
\usepackage{graphicx}
\usepackage{float}
\usepackage{subcaption}
 
\usepackage{multirow}
\usepackage{siunitx}

\newcommand{\bea}{\begin{eqnarray*}}
\newcommand{\eea}{\end{eqnarray*}}
\newcommand{\beao}{\begin{eqnarray}}
\newcommand{\eeao}{\end{eqnarray}}

\Title{A Lightweight Algorithm to Model  Radiation Damage Effects in Monte Carlo Events for High-Luminosity Large Hadron Collider Experiments
}

\TitleCitation{A Lightweight Algorithm to Model  Radiation Damage Effects in Monte Carlo Events for High-Luminosity Large Hadron Collider Experiments}

\Author{{Keerthi Nakkalil} 
 \orcidA{}  and Marco Bomben *\orcidB{}}

\AuthorNames{Keerthi Nakkalil and Marco Bomben}

\AuthorCitation{{Nakkalil, K.;} 
 Bomben, M.}

\address[1]{%
{APC,} 
 Université Paris Cité, CNRS/IN2P3, {75013} 
 Paris, France; keerthi.nakkalil@cern.ch\\
}

\corres{\hangafter=1 \hangindent=1.05em \hspace{-0.82em} Correspondence: marco.bomben@cern.ch}

\abstract{Radiation damage significantly impacts the performance of silicon tracking detectors in Large Hadron Collider (LHC) experiments such as ATLAS and CMS, with signal reduction being the most critical effect; adjusting sensor bias voltage and detection thresholds can help mitigate these effects, generating simulated data that accurately mirror the performance evolution with the accumulation of luminosity, hence fluence, is crucial
	.
 The ATLAS and CMS collaborations have developed and implemented algorithms to correct simulated Monte Carlo (MC) events for radiation damage effects, achieving impressive agreement between collision data and simulated events.  
In preparation for the high-luminosity phase (HL-LHC), the demand for a faster ATLAS MC production algorithm becomes imperative due to escalating collision, events, tracks, and particle hit rates, imposing stringent constraints on available computing resources. This article outlines the philosophy behind the new algorithm, its implementation strategy, and the essential components involved.  The results from closure tests indicate that the events simulated using the new algorithm agree with fully simulated 
events at the level of few \%.  The first tests on computing performance show that the new algorithm is as fast as it is when no radiation 
damage corrections are applied.}

\keyword{silicon radiation detectors; simulations; radiation damage} 

\begin{document}


\section{Introduction}
\label{sec:intro}
Silicon radiation detectors are at the core of experiments conducted at high-energy colliders, such as the CERN Large Hadron Collider (LHC) \cite{LHC}. Multiple layers of pixel and strip detectors are utilized to accurately measure the momentum and vertices of charged particles. Examples of such detector systems are part of the CMS \cite{CMScollaboration} Silicon Tracker \cite{Karimaki:368412} and the ATLAS \cite{AtlasDetector} Inner Detector \cite{Haywood:331064}. 

Hybrid pixel detectors, positioned close to the beam's interaction point, offer fine segmentation of a few tens of micrometers, ensuring spatial resolution on a single point down to \SI{10}{\micro\meter}. Precision measurements and the discovery potential heavily rely on the performance of tracking devices, particularly for energy/momentum resolution and flavor identification. Notably, both ATLAS \cite{IBL_paper,AtlasPixels} and CMS \cite{CMSPixels} pixel detectors are equipped with thin planar n$^+$-on-n silicon sensors {(ATLAS Innermost B-Layer includes sensors made in 3D n$^+$-on-p technology in the forward region).} 

The LHC is scheduled for an upgrade to a high-luminosity machine (HL-LHC) \cite{Apollinari:2284929}, with the goal of integrating a dataset of 4000 fb$^{-1}$; this is twenty times larger than its current capacity, allowing for the full exploitation of the physics reach of the LHC machine. To integrate such a large dataset within a reasonable time, the instantaneous luminosity will need to increase by a factor of 5--7 beyond the nominal value, reaching 7 $\times 10^{34}$ cm$^{-2}$ s$^{-1}$. This increase in instantaneous luminosity will lead to a corresponding rise in particle flux, track densities, and hit rates. Consequently, the current inner tracking systems will need to be replaced to cope with this challenging environment, as radiation levels affecting both sensors and electronics will also increase by a similar amount.

The effects of radiation damage are already measurable at the fluences of LHC 
Run 2 and Run 3. For instance, the charge collection efficiency (CCE) of the innermost ATLAS Pixel layer has now reduced  to approximately $\sim $70\%  at a fluence of about $1\times
10^{15}$ \si{n_{eq} \cm^{-2}} \cite{EPS2023Bomben}. 

The ATLAS, CMS, and LHCb collaborations have developed algorithms to incorporate radiation damage effects into their Monte Carlo (MC)-simulated events \cite{Dawson:2764325}. Thanks to these algorithms, it is possible to track the evolution of pixel detector performance as radiation damage accumulates. Despite the different approaches adopted by the three collaborations, the level of agreement between data and Monte Carlo (MC) simulations is remarkable, achieving agreement at the $O(\%)$ level \cite{EPS2023Bomben}. These algorithms enable the prediction of operational parameters such as depletion voltage and make possible the generation of samples for future detector conditions; these are essential for training reconstruction algorithms.

Despite the importance and success of these algorithms, some may not be viable for the HL-LHC phase due to computing resource constraints. This is particularly true for the ATLAS Pixel detectors. Therefore, there is a pressing need for faster yet equally precise algorithms to meet the demands of the HL-LHC data-taking conditions.

In this paper, a new lightweight algorithm is presented which is able to 
address the challenges posed by the HL-LHC on computing resources for 
the inclusion of radiation damage effects in MC simulated events. 
The paper is organized as follows. 

In Section \ref{sec:digi}, the ATLAS  algorithm for simulating radiation damage effects to silicon pixel sensors is presented along with some key results. Section \ref{sec:lut} introduces the new algorithm, which relies on Look-Up Tables (LUTs). This section also discusses the various types of simulations---TCAD {(Technology--Computer-Aided Design)} and Geant4 \cite{Geant41,Geant42,Geant43} based---that serve as inputs to the new algorithm.

In order to validate the new algorithm, a series of closure tests were conducted, comparing events simulated using the LUTs algorithm to fully simulated events. The results of these closures tests 
are reported in Section \ref{sec:closure}. Finally, the paper concludes (Section \ref{sec:conclusions}) with a discussion on the results and an outlook for the project.


\section{The ATLAS IBL and Pixel Radiation Damage Digitizer}
\label{sec:digi}

The ATLAS Collaboration developed and implemented an algorithm to  include the effects of radiation damage to silicon pixel sensors in the 
Monte Carlo-simulated events. In the following (Section \ref{sec:ATLASPixels}), the ATLAS Pixel and IBL detectors will be briefly discussed. The radiation damage digitizer for these detectors will be presented in Section \ref{sec:Run3digi}.

\subsection{The ATLAS IBL and Pixel Detector}
\label{sec:ATLASPixels}

The ATLAS Pixel detector \cite{AtlasPixels} has been in operation since the beginning of LHC Run1. It comprises three barrel layers and three disks on each side. The barrel layer closest to the interaction point (IP) is called the B-Layer and is positioned at a radius $R = $ 50.5 mm from the IP. All layers are composed of hybrid pixel modules, with sensors that are planar n$^{+}$-on-n, 250 \si{\micro\meter} thick, and have a pitch of 50 $\times$ 400 \si{\micro\meter}$^2$ (bending and transverse planes, respectively).

For Run 2 of the LHC, which started in 2015, a new pixel layer was added: the Insertable B-Layer (IBL) \cite{IBL_paper}, positioned \SI{33}{\milli\meter} from the IP. The IBL is also composed of hybrid pixel modules. Compared to those in the Pixel detector, IBL modules feature thinner sensors, 200 \si{\micro\meter} instead of 250 \si{\micro\meter}, and a finer pitch of 50 $\times$ 250 \si{\micro\meter}$^2$. Planar n$^{+}$-on-n technology is used for IBL sensors, except in the forward region, where novel 3D n$^{+}$-on-p sensors are employed.


By the end of 2023, the IBL (B-Layer) planar sensors had integrated a fluence of approximately 1.2 (1.0) $\times 10^{15}$ \si{n_{eq} \cm^{-2}}. These fluence levels led to a signal loss of about $30\%$ for both detectors.

\subsection{The Radiation Damage Digitizer for Pixel and IBL Detectors}
\label{sec:Run3digi}

Signal reduction is the most important radiation damage effect for the performance of silicon pixel tracking detectors in ATLAS. It is important to have simulated data that reproduce the evolution of performance with the increase in radiation fluence. ATLAS collaboration developed and implemented an algorithm that reproduces signal loss and changes in Lorentz angle due to radiation damage \cite{RadDamageDigi_2019}. This algorithm is now the default for Run 3-simulated events \cite{ATL-SOFT-PUB-2021-001}.

The algorithm uses electric field profiles produced using TCAD tools  to evaluate the time carriers will take to reach the collecting electrode. 
The electric field is deformed by the presence of defects due to radiation  damage; in particular, it has a deep minimum in the sensor mid-plane \cite{bib:DP}.  This has two main effects: first and most important, carriers will have  reduced velocity and so more prone to be trapped; second, the Lorentz  angle will become dependent on the position along the bulk.  The combination of these effects determines a reduction in the CCE,  a deformation of the cluster shape, and, eventually, a degradation of  the space--point resolution. 

\begin{figure}[H]
\includegraphics[width=0.7\textwidth]{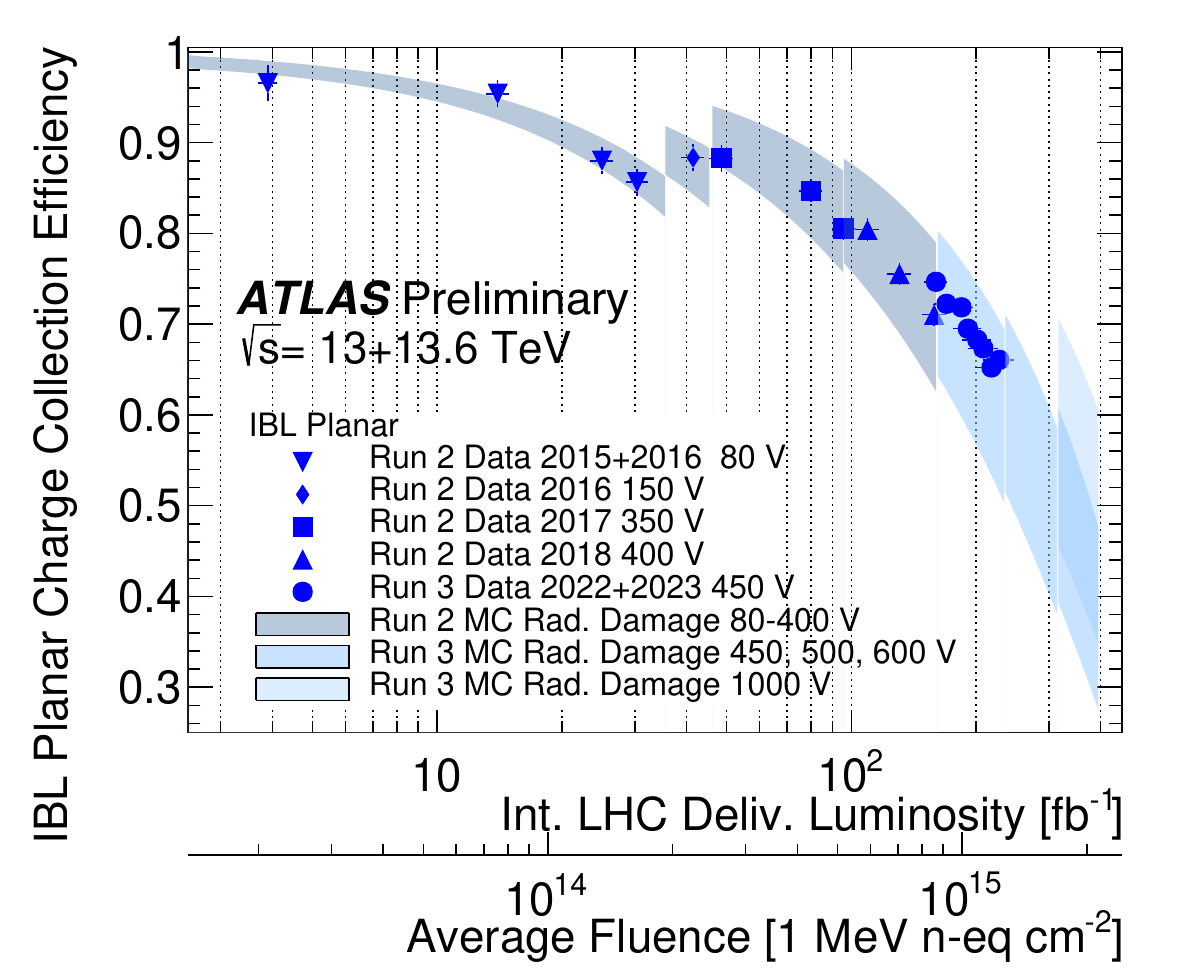}
\caption{\label{fig:IBLPL-CCEVsIntLum-MCrad-Prel}{Comparison} 
 of the evolution of CCE with integrated luminosity for IBL  planar sensors in data (points) and MC events (bands) \cite{Battaglia:2866558}; the corresponding radiation fluence is also indicated. The vertical bands include the estimated uncertainty affecting the input parameters of TCAD radiation damage model and of the trapping constants.}
\end{figure}

In Figure \ref{fig:IBLPL-CCEVsIntLum-MCrad-Prel}, a comparison  between the data and the simulated MC events for the CCE in IBL planar sensors is shown. 

The comparison covers almost two order of magnitude of fluence and  the level of agreement is excellent, down to 1 \%. More results can be found in \cite{EPS2023Bomben}. 

Despite the success of this algorithm, it cannot be used for the HL-LHC as it is excessively time-consuming. Therefore, a new, faster, yet equally precise algorithm is required, which will be presented in the next section.


\section{A Radiation Damage Digitizer Based on Look-Up Tables for Silicon Detectors at  HL-LHC Experiments}
\label{sec:lut}

Starting in 2025 and extending through 2028, the LHC will be upgraded to the high-luminosity LHC (HL-LHC), making the expansion of experimental particle physics research and the exploration of fundamental physics of the LHC possible. The HL-LHC will deliver an instantaneous luminosity of $7.5 \times 10^{34}$ \si{\cm^{-2}s^{-1}}, a 5--7 times increase from the nominal value resulting in an average pileup of about 200 inelastic proton--proton collisions per bunch crossing. The innermost parts of the pixel detector, situated closest to the interaction point, will be exposed to unprecedented levels of radiation, reaching a non-ionizing radiation fluence of $1-2\times 10^{16}$ \si{n_{eq} \cm^{-2}}  during the expected lifetime of the ATLAS detector \cite{CERN-LHCC-2017-021,ITkStripsTDR}. 

\begin{figure}[H]
  \centering
  \begin{subfigure}{0.43\textwidth}
    \includegraphics[width=\linewidth]{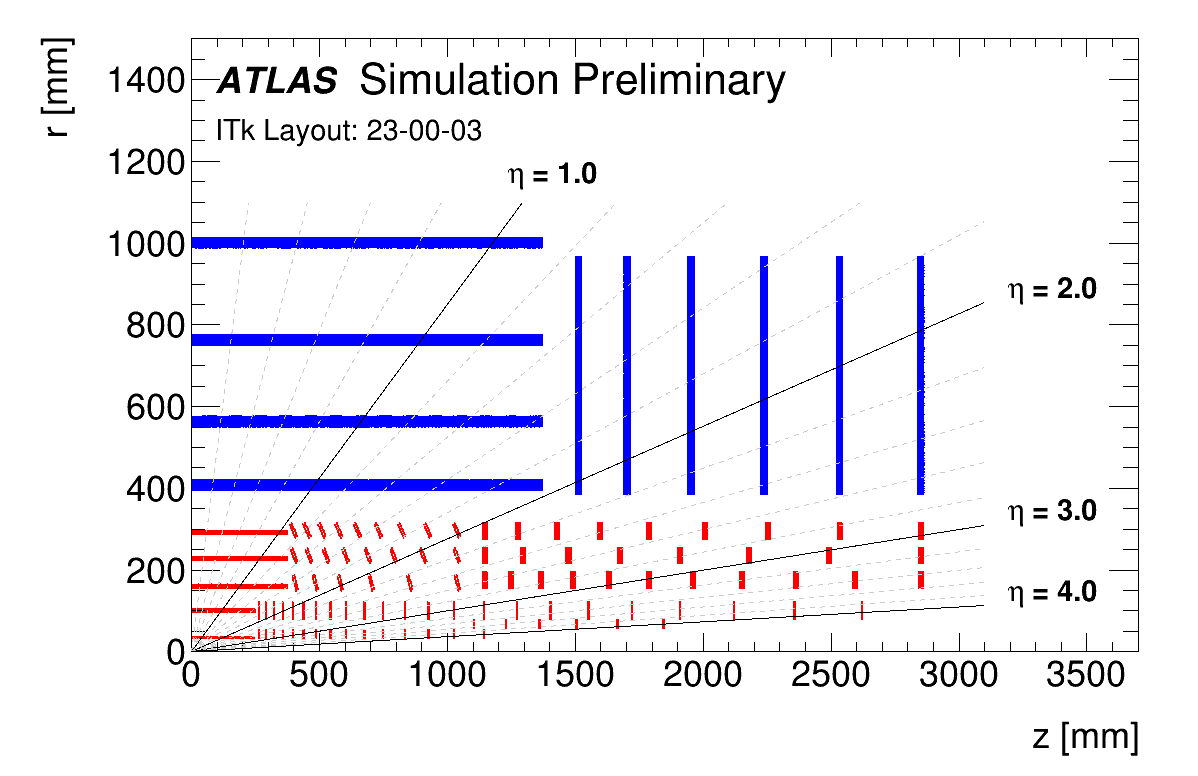}
    \caption{}
    \label{fig:itk_layout}
  \end{subfigure}
  \hfill
  \begin{subfigure}{0.56\textwidth}
    \includegraphics[width=\linewidth]{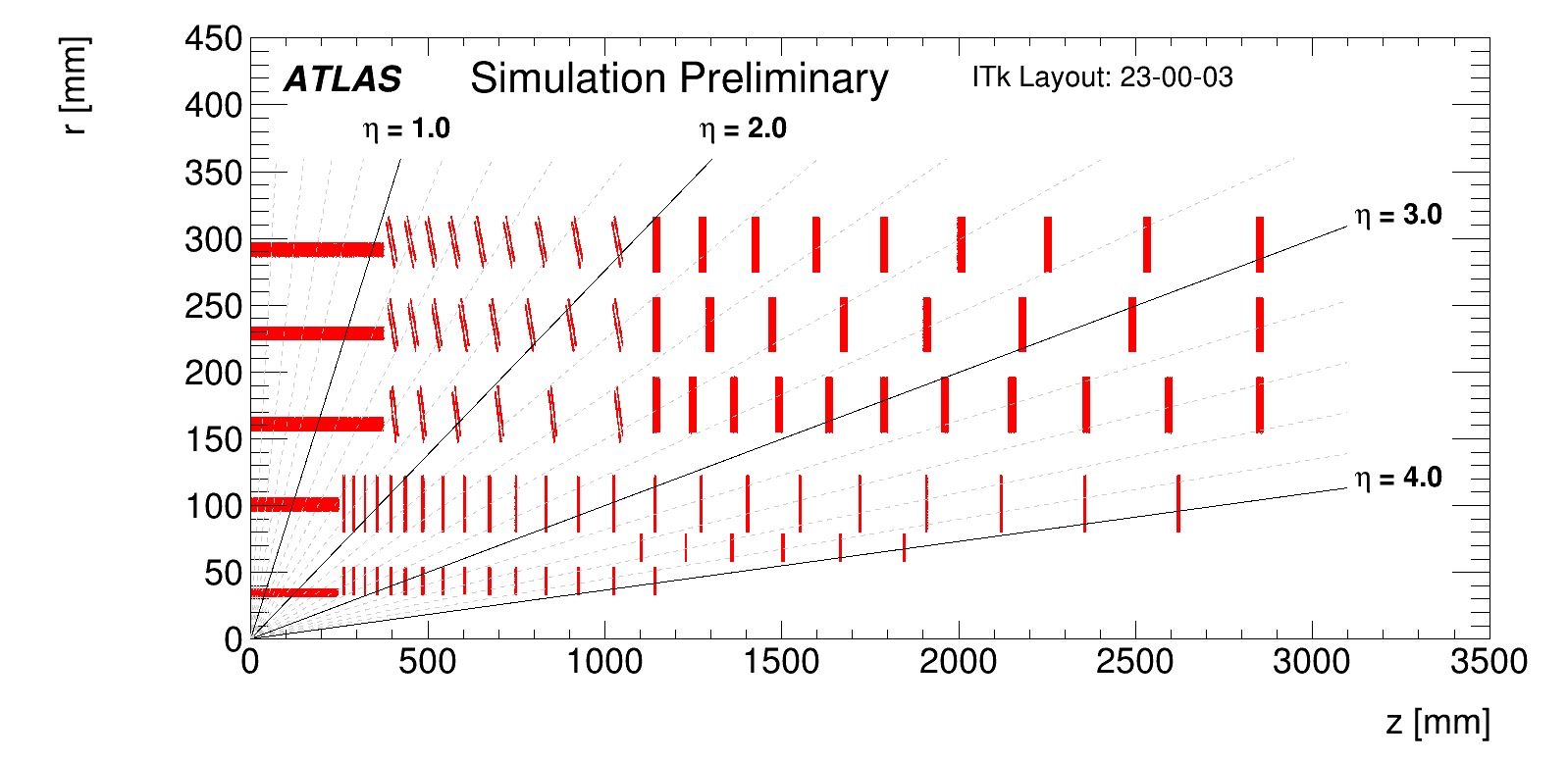}
    \caption{}
    \label{fig:itk_pixel_layout}
  \end{subfigure} 
 \caption{\label{fig:ITk}ATLAS ITk layout. (\textbf{a}) Full detector; pixels/strips parts are in red/blue. (\textbf{b}) Pixel part. The $z$ coordinate is measured along the beam axis and $r$ is the radial distance from the center of ATLAS, from \cite{ATL-PHYS-PUB-2021-024}.}
\end{figure}

The actual ATLAS IBL, Pixel, and Strip detectors will not be able to cope with such harsh conditions. For this reason, the current detectors will be replaced by the  ATLAS Inner Tracker (ITk), a  full-silicon vertexing and tracking detector \cite{CERN-LHCC-2017-021,ITkStripsTDR}. The ATLAS ITk  will have pixels detectors in its innermost part and strips at  the outer radii. The layout of ATLAS ITk and that of its pixels part are presented in Figure \ref{fig:ITk}.

The ATLAS ITk pixels detector comprises five barrel layers (from L0 to L4) and a system of as many rings covering the forward region.
The innermost pixel layer (L0) and ring (R0) will be equipped with  3D n$^+$-on-p sensors. All the rest of the detector will be equipped with 
planar n$^+$-on-p sensors, with thicknesses of either \SI{100}{\micro\meter} (in L1 and R1) or \SI{150}{\micro\meter} (in all other layers and rings).

The radiation damage fluences expected after 2000 (4000) fb$^-1$  of integrated luminosity by the sensors in L0 (L1) are 18(4) $\times 10^{15}$ \si{n_{eq} \cm^{-2}}, including the safety factors.

The anticipated increase in particle density at HL-LHC calls for a faster algorithm to model the effects of radiation damage in the ATLAS MC events. A novel methodology centered on charge re-weighting from Look-Up Tables (LUTs) is under study and is outlined in the subsequent sections.

\subsection{LUT Method}
\label{ssec:LUT_method}


\textls[-25]{Similar to the actual radiation damage digitizer (see Section \ref{sec:Run3digi}), the Look-Up Table (LUT) method simulates radiation damage effects in ATLAS MC events after Geant4  \cite{Geant41,Geant42,Geant43} simulates charge deposition and before signal digitization, where signal is discriminated and transformed into digital format by simulating the response of the readout electronics chip.}

The task of the radiation damage digitizer is to estimate  on which pixels the deposited charges will induce a signal and how large this signal will be. The LUT method is proposed to accomplish this task for the data taking of ATLAS during the high-luminosity phase 
of LHC (HL-LHC). ATLAS MC events will be simulated assuming an ideal, undamaged pixel sensor. The impact of radiation damage will then be incorporated as a correction, utilizing the LUTs.

The LUT method presented herein is inspired by the ``template'' method used by the CMS collaboration \cite{Swartz:2008oU}. The primary distinction between the CMS and LUT methods is that, while the former applies corrections to simulated clusters of pixels, the latter aims to reweigh observables of a group of drifting carriers in the silicon bulk.

\begin{figure}[H]
  \includegraphics[width=1.0\textwidth]{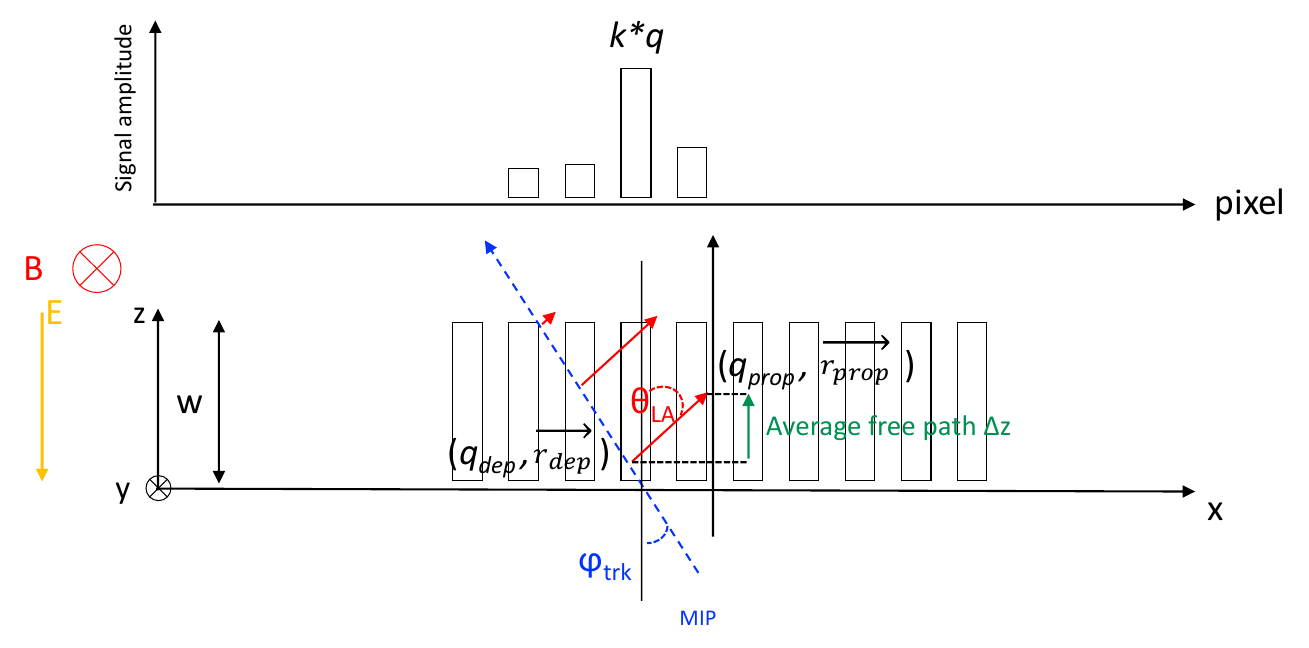}
  \caption{\label{fig:MIPS}A schematic diagram of carrier dynamics in silicon planar pixel sensors. As a MIP crosses the sensor (at 
  an angle $\varphi_{trk}$), electron--hole pairs are created and transported to their respective electrodes under the influence of electric and magnetic fields. Electrons and holes may be trapped before reaching the electrodes, but still induce a charge on the primary and neighbor electrodes.} 
\end{figure}

The design of the  LUT method to model the radiation damage effects in ATLAS MC events 
is based on the dynamics of charge carriers in the bulk of pixel sensors. The situation is sketched 
in Figure \ref{fig:MIPS} for the case of planar sensors.

Here, the sensor is assumed to be a planar pixel sensor of thickness $w$; $y$ is the  direction of the magnetic field $\vec{B}$ and the direction of the  electric field  $\vec{E}$ is along the bulk depth axis $z$; carriers are deflected  in the $x$ direction due to magnetic field.

Electrons and holes produced by a minimum ionizing particle (MIP) traversing the sensor drift toward collecting electrodes at an angle known as the Lorentz angle ($\theta_{LA}$) with respect to the electric field. 

In pristine detector---i.e., without radiation damage effects---the final position of the carriers is calculated 
projecting the carrier position to the respective collecting  electrode. The final position is then smeared to take into account diffusion. The LUT method is intended to apply corrections to this basic method. In the following the initial position of the charge carrier will be referred to as {\it {deposited}} position $\vec{r}_{dep}=(x_{dep},y_{dep},z_{dep})$, and the final one the {\it {propagated}} position $\vec{r}_{prop}=(x_{prop},y_{prop},z_{prop})$; see also 
Figure \ref{fig:MIPS}.
 
In Figure \ref{fig:MIPS}, drifting carriers can be trapped before reaching the collecting electrode due to radiation damage in the sensor bulk. In such cases, they induce a signal that is only a fraction, $k<1$, of their charge. This fraction can be calculated using the Shockley--Ramo theorem \cite{ShockleyPot,Ramo}, which requires the initial and final positions of the carrier as input. 

The carriers that are trapped cover a distance $\Delta z = z_{prop} - z_{dep}$ along the bulk depth direction, which depends on the $\vec{r}_{dep}$ position of the carriers. This dependence on the carrier generation position is due to the non-uniform electric field in the sensor bulk after radiation damage \cite{EPS2023Bomben}. As mentioned above, the carriers will drift at an angle $\theta_{LA}$ relative to the direction of the electric field. This angle also depends on the $\vec{r}_{dep}$ position where the carrier was created, again due to the non-uniformity of the electric field. The carrier will end up at a position in the bending direction (orthogonal to the electric and magnetic field) that is displaced---on average, (the carrier movement is affected by diffusion too which adds dispersion to the final carrier position)---by $\Delta x = \tan (\theta_{LA}) \cdot \Delta z$; of course, $x_{prop} = x_{dep} + \Delta x$, if there was no diffusion. As discussed above, the carrier's final position $\vec{r}_{prop}$ will determine the amplitude of the signal induced on the electrodes. This final position depends on the carrier's initial position $\vec{r}_{dep}$ and on the combination of electric and magnetic fields. The signal will be a fraction $k$ of the carrier charge $q$, and this fraction $k$ is also a function of the original carrier position $\vec{r}_{dep}$.

In summary, for each group of carriers, the method calculates the free path $\Delta z$, the Lorentz angle $\theta_{LA}$, and the fraction of the induced signal $k$. These three quantities will be referred to as the {\it {observables}}.

To determine the values of the three observables, repeated simulations of the drift of carriers deposited at precise positions $\vec{r}_{dep}$ within the bulk are conducted. These simulations are performed using the  Allpix$^2$ \cite{APSQPaper} simulation framework, which will be presented in Section \ref{ssec:Allpix}, using precise electric field and Ramo potential maps produced 
using TCAD {(Technology--Computer-Aided Design)} simulations.

For each simulated event, the initial position $\vec{r}_{dep}$,   the final position $\vec{r}_{prop}$  of the carrier will be saved, together with the 
signal fraction $k$ induced on the pixel matrix. In principle, the values of three observables should be recorded as a function of the initial position $\vec{r}_{dep}$
of the carrier, but due to the aforementioned limited  computing resources, the LUTs will be a function of only  $z \equiv z_{dep}$. 
For a 
fixed $z$ value, the average over all $(x_{dep},y_{dep})\equiv(x,y)$  positions will be carried out for each 
observable and assigned to that $z$ value; to make the notation lighter, ``{\it dep}'' is dropped from here onward.
Thus, for a single event and a single value of $z$, the three LUTs will be:
\begin{equation}
    \begin{cases}
    \Delta z(z)|_{1}  & =   \sum_{x,y}  \Delta z(x,y,z)/ \sum_{x,y}\\
    \theta_{LA}(z)|_{1} & =   \sum_{x,y}\theta_{LA}(x,y,z)/\sum_{x,y}  \\ 
    k(z)|_{1} & =   \sum_{x,y}  k(x,y,z)/\sum_{x,y}  \label{eq:LUTs} \\
\end{cases}
\end{equation}

Using the LUTs defined in \ref{eq:LUTs}, the propagated position $\vec{r}_{prop}$  
and the induced signal $q_{ind}$ of a charge, $q$ deposited at depth $z$ in the sensor bulk is calculated as follows: 

\begin{equation}
\begin{cases}
    x_{prop} & =  x + [\tan(\theta_{LA}(z) )\cdot \Delta z(z)]  + \Delta^{diff}_{x} \\
    y_{prop} & =  y + \Delta^{diff}_{y}  \\ 
    z_{prop} & =  z + \Delta z(z) \\
    q_{ind}  & =  k(z)\cdot q\label{eq:rprop}
\end{cases}
\end{equation}
where $\Delta^{diff}_{x,y}$ is a Gaussian-distributed random number added to simulate the effect of diffusion. 

The essence of the LUT method is summarized in Equation (\ref{eq:rprop}): the precise dynamics of carrier drift are substituted with an ``average'' drift, and the same principle applies to the signal amplitude.

The procedure of charge deposition and drifting is repeated  for each $\vec{r}_{dep}$ several times in order to assess the dispersion of the carriers dynamics.

It is worth noting that, while this study primarily focuses on planar pixel sensors, the same methodology can also be readily applied to strip sensors. Additionally, for the sake of simplicity, the method will be presented here only for planar pixel sensors, but ongoing efforts are being made to extend it to 3D sensors as well---see, for example \cite{EPS2023Bomben}.

In the following sections, the Allpix$^2$ simulation framework, the TCAD simulations utilized as input to Allpix$^2$ simulations, and the process of calculating the Look-Up Tables (LUTs) will be presented.

\subsection{Allpix-Squared for Radiation Damage Digitizer}
\label{ssec:Allpix}

Allpix$^{2}$ is a generic, open-source software framework for the simulation of silicon pixel detectors \cite{APSQPaper}. The framework allows the user to create detailed simulations of the entire experimental chain of a testbeam, from incident radiation to digitized detector response. An extensible system of modules is responsible for executing the distinct simulation steps, such as realistic charge carrier deposition using the Geant4 toolkit and the propagation of charge carriers in silicon through a drift--diffusion model. Detailed electric field maps imported from TCAD simulations can be used to model the drift behavior of charge carriers within the silicon, introducing a higher level of realism to Monte Carlo-based simulations of particle detectors. 

\begin{figure}[H]
  \includegraphics[width=0.9\textwidth]{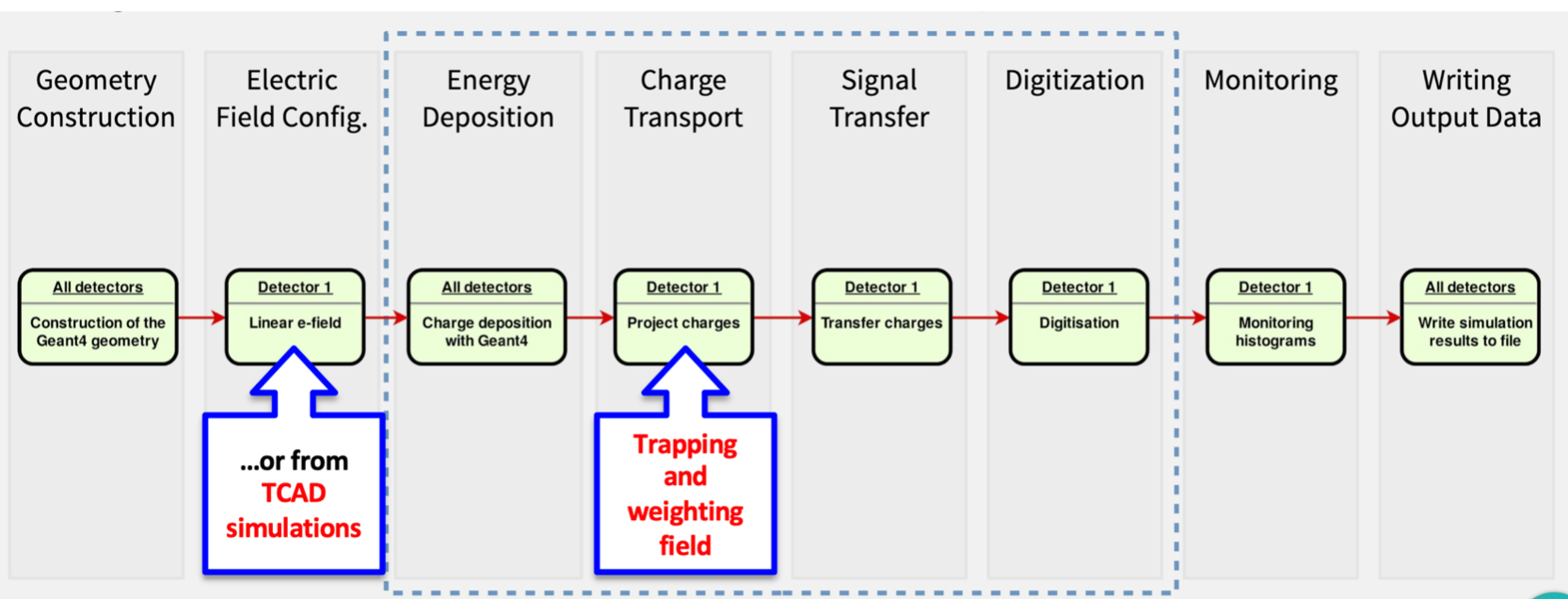}
  \caption{{Modified} 
 Allpix$^{2}$ simulation chain with a single detector for radiation damage digitizer.}
  \label{fig:APSQ_SimChain_RadDam}
\end{figure}

In this study, a planar n$^+$-on-p pixel sensor with a thickness of \SI{100}{\micro\meter} and a pitch of $50 \times 50$ \si{\micro\meter\squared} is simulated. This type of sensor is representative of what will be used in the second-to-innermost pixel layer (L1) of the ITk pixel detector. The sensor is simulated after irradiation corresponding to a fluence of $4 \times 10^{15}$ \si{n_{eq} \cm^{-2}} and operated at a voltage of \SI{600}{\volt}. These conditions are expected for the aforementioned pixel layer at the end of the HL-LHC phase of ATLAS data taking. Figure \ref{fig:APSQ_SimChain_RadDam} illustrates the simulation chain implemented for evaluating the LUTs, highlighting the use of electric field and Ramo potential maps from TCAD, along with the simulation of trapping effects. 

\subsubsection{Simulation Inputs}
\label{sssec:SimulationInputs}

The electric field and Ramo potential maps calculated using TCAD tools can be inputted into Allpix$^2$ simulations to model the behavior of silicon radiation detectors, as outlined in Section \ref{ssec:LUT_method}. The following sections provide details of the TCAD simulations used for the calculation of LUTs in this study. All simulation results presented here have been obtained 
using Silvaco {TCAD} (\url{https://silvaco.com/tcad/}, {accessed on 17 June 2024)} 
tools.

\begin{figure}[H]
  \centering
  \begin{subfigure}{0.35\textwidth}
    \includegraphics[width=\linewidth]{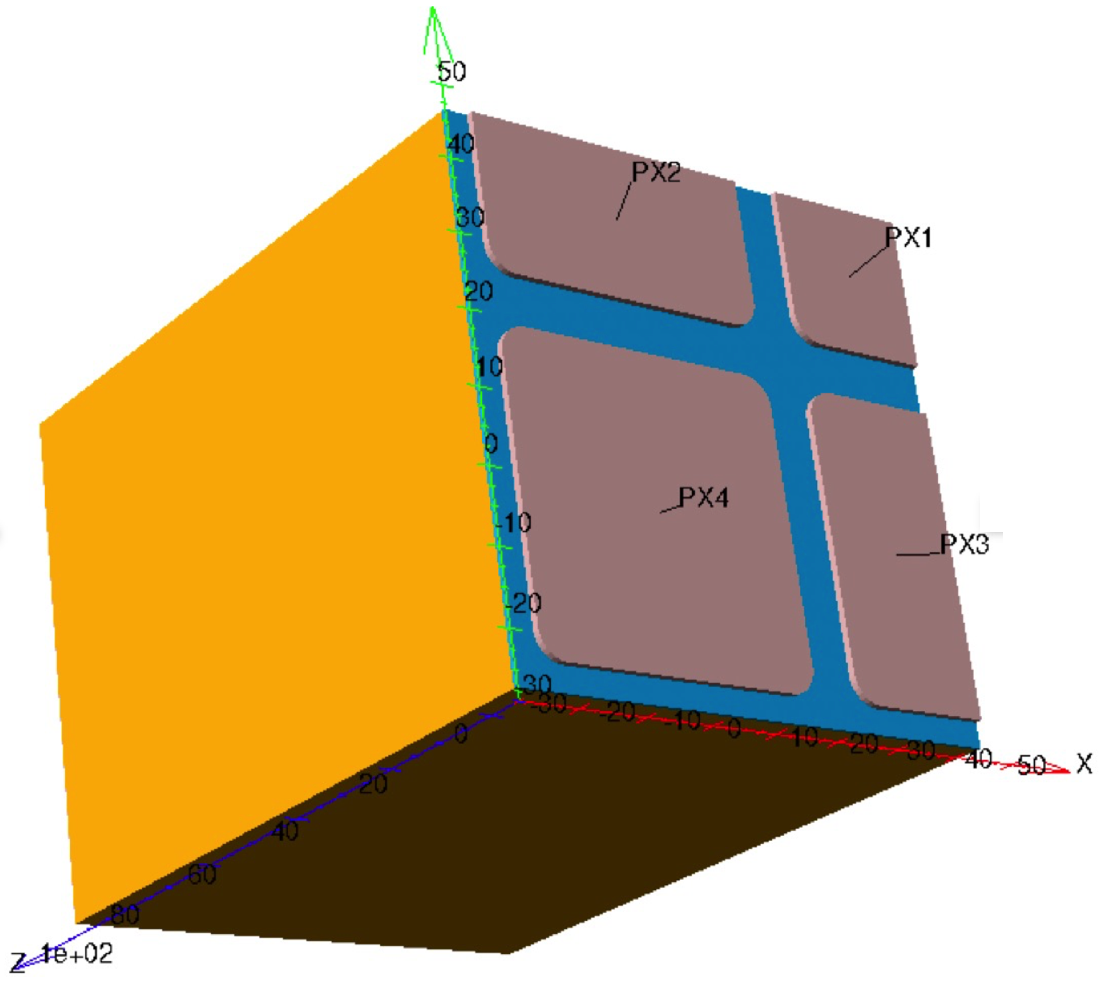}
    \caption{}
    \label{fig:TCAD_structure}
  \end{subfigure}
  \hfill
  \begin{subfigure}{0.35\textwidth}
  \hskip-25mm  \includegraphics[width=\linewidth]{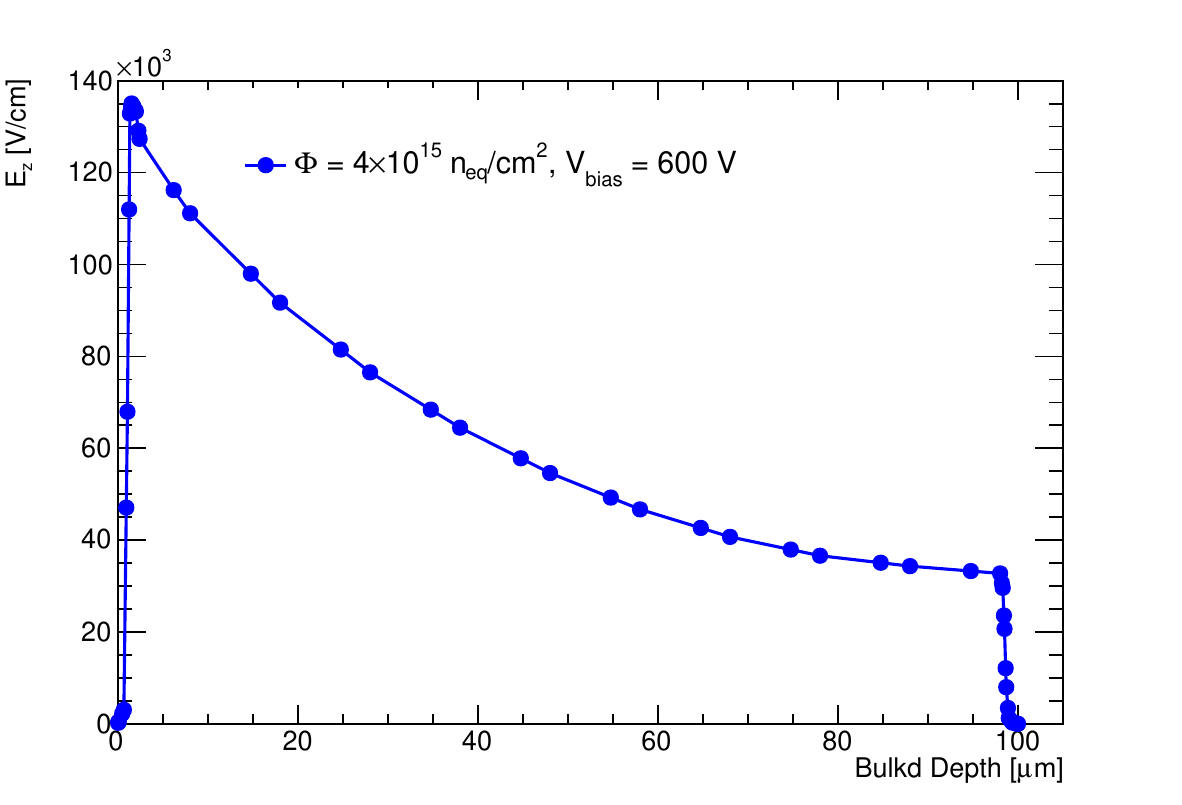}
 \caption{}
    \label{fig:TCAD_Efield}
  \end{subfigure}\\
\begin{subfigure}{0.35\textwidth}\vspace{+6pt}
  \hskip-15mm  \includegraphics[width=\linewidth]{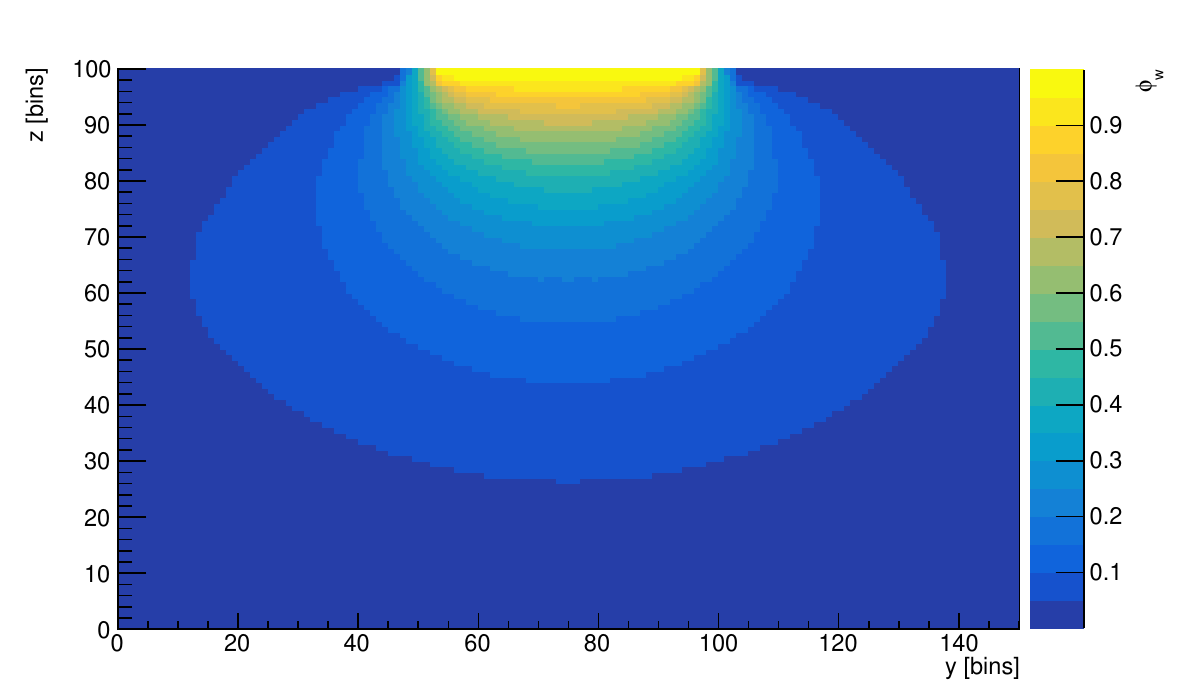}
        \caption{}
    \label{fig:TCAD_Ramo}
  \end{subfigure}
 \caption{ \label{fig:overall}{(\textbf{a}) Depicts} 
one-quarter of a $ 3 \times 3$ pixel matrix structure simulated in TCAD and (\textbf{b}) illustrates the extracted electric field distribution as a function of charge generation depth of the structure in (\textbf{a}) irradiated at a fluence of $4 \times 10 ^{15}$ \si{n_{eq}\cm^{-2}} at \SI{600}{\volt} bias voltage. (\textbf{c}) A 2D projection of the 
 weighting potential for the simulated structure.}
\end{figure}

The structure simulated in TCAD represents one-quarter of a $3 \times 3$ pixels matrix from a \SI{100}{\micro\meter} n$^+$-on-p planar pixel sensor with a pitch of $50 \times \SI{50}{\micro\meter}^2$; refer to Figure \ref{fig:overall}a,c. The simulation of one pixel plus its neighbors is required to model charge sharing accurately via Ramo potential calculation. However, due to the symmetry of the design, it is sufficient to simulate just one-quarter of a $3 \times 3$ pixels matrix.

To simulate the effects of radiation damage, the LHCb TCAD radiation damage model \cite{LHCb_TCAD_RadDam} was used. This model was developed for planar n$^+$-on-p pixels intended for the LHCb Velo Upgrade \cite{Collaboration:1624070}, where irradiation fluences of several $1 \times 10^{15}$ \si{n_{eq} \cm^{-2}} are expected. Similarly, planar n$^+$-on-p pixel sensors in the ATLAS ITk detector will be exposed to similar fluences during their expected operational lifetime. Therefore, the LHCb TCAD radiation damage model was chosen for this study.

In the ATLAS ITk detector, planar pixels will be utilized everywhere except in the innermost section of the pixel detector, where 3D sensors will be employed \cite{CERN-LHCC-2017-021}.

The largest fluence expected for planar pixels is $\Phi_{\text{max}} \sim 4 \times 10^{15}$~\si{n_{eq} \cm^{-2}}~\cite{CalderiniVertex2021}. Electric field maps were obtained for a sensor irradiated at the fluence $\Phi_{\text{max}}$ once polarized at the maximal expected voltage (600 V). The extracted electric field profile along the bulk depth is shown in \cref{fig:overall}a,b. In Figure \ref{fig:overall}a,c 2D projection of the 
weighting field is depicted; here, the result is presented for the full 3 $\times$ 3 pixels matrix as it is already in the format for Allpix$^{2}$.

The electric field and weighting potential maps from TCAD are integrated into the Allpix$^{2}$ simulation chain, as illustrated in \cref{fig:APSQ_SimChain_RadDam}, to generate the three LUTs introduced in \cref{ssec:LUT_method} for correcting pixels response in ATLAS simulations. In the following section, the 
evaluation of LUTs using Allpix$^2$ will be presented.

\subsubsection{Look-Up Tables Generation}
\label{sssec:GeneratingLUTs}

Charged particles produced in LHC collisions will ionize the  pixels sensors volume homogeneously. So in order to obtain a  realistic representation of the pixel response, it is important  to study it for all possible charge deposition locations $\vec{r}_{dep}$.
The LUTs are generated using the ``scan'' mode of the ``DepositionPointCharge'' module in Allpix$^2$ \cite{APSQPaper}. This module places a specified quantity of charge carriers at a precise location within the detector's active volume. 
In the studies reported here, the scan mode  was configured to deposit $q_{dep} = $ 1000 electron--hole pairs every $\SI{2}{\micro\meter}$ along the $z$ direction and every $\SI{1}{\micro\meter}$ for the other two directions for the simulated device. 

\begin{figure}[H]
  \includegraphics[width=0.45\textwidth]{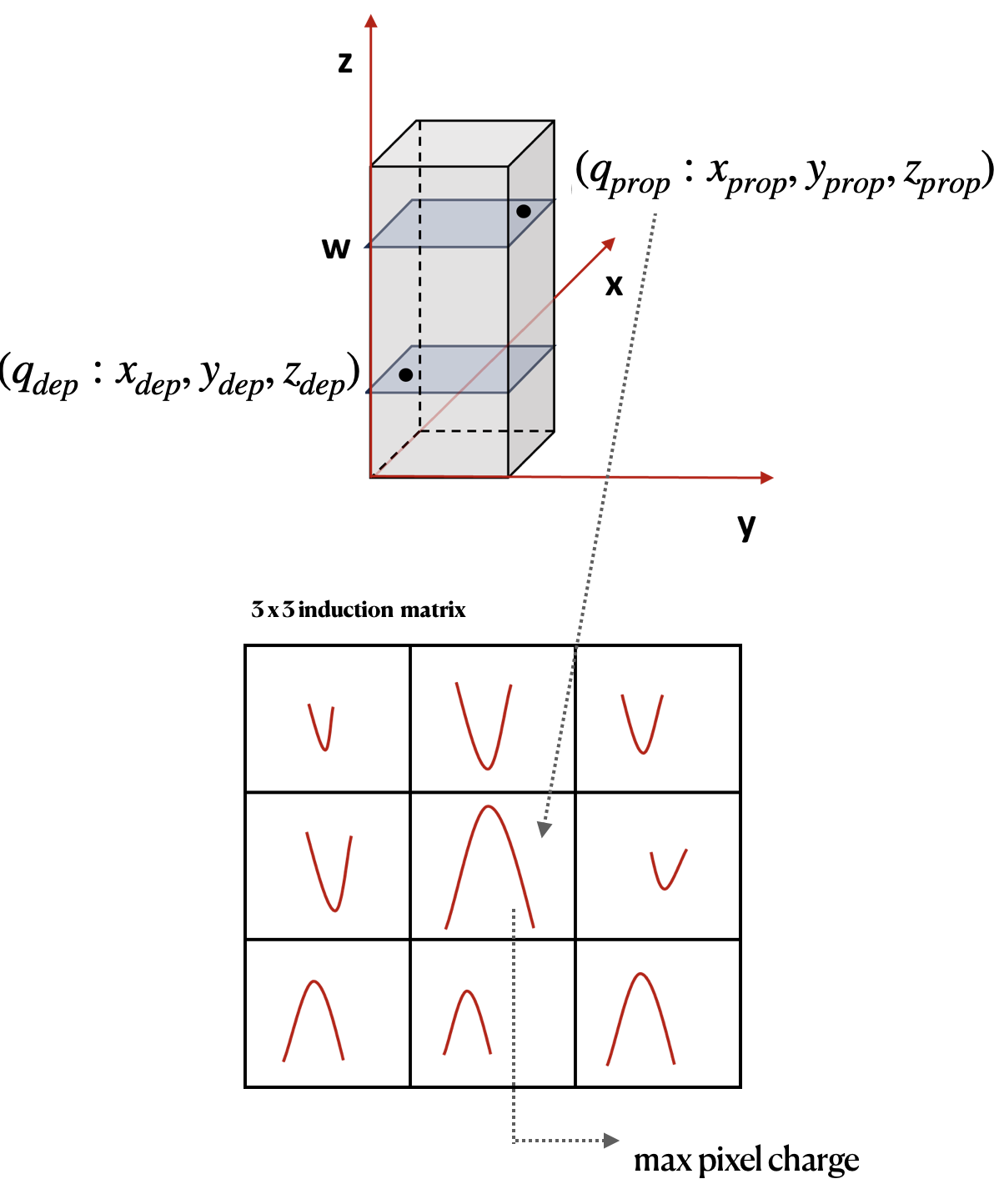}
  \caption{(top) Illustration of deposited (propagated) charge $q_{dep}(q_{prop})$ and position $\vec{r}_{dep}(\vec{r}_{proop})$. (bottom) Illustration of the 3 $\times$ 3  matrix used to evaluate the induced signal; the central pixel is the one containing $q_{prop}$; hence, $\vec{r}_{prop}$.}
  \label{fig:PixMaxCharge}
\end{figure}

Once carriers have drifted to their final position $\vec{r}_{prop}$, the induced signal is evaluated for the pixel cell where they ended up and for the first eight neighbors, as illustrated in Figure \ref{fig:PixMaxCharge}. It is important to note that, while the propagated charge may land in the same pixel cell as the deposited charge, this is not always the case.

In the following, the propagated position will always correspond to that of the electrons. This assumption is made in calculating the LUTs due to the negligible impact of holes on signal ``position'' compared to electrons. This simplification is based on the principle known as the ``small-pixel effect'' \cite{rossi2006pixel}.
In heavily segmented sensors such as pixel sensors, the weighting potential profile along the bulk has a steep gradient near the readout side. Consequently, only carriers in close proximity to the pixel electrode can induce a significant signal. Given that holes move away from the readout electrode, their influence on signal position can be safely  neglected.

The pixel with the largest induced signal is identified,  its position and signal $q_{max}$ are retained for the subsequent steps.
The CCE per deposition position $\vec{r}_{dep}$ is defined as $q_{max}$ divided by the deposited charge $q_{dep}$:
\begin{equation}
    CCE(\vec{r}_{dep}) = q_{max}/q_{dep}
    \label{eq:CCE_r_dep}
\end{equation}

\begin{figure}[H]
\centering
\begin{subfigure}{0.49\textwidth}
\centering
\includegraphics[width=0.9\textwidth]{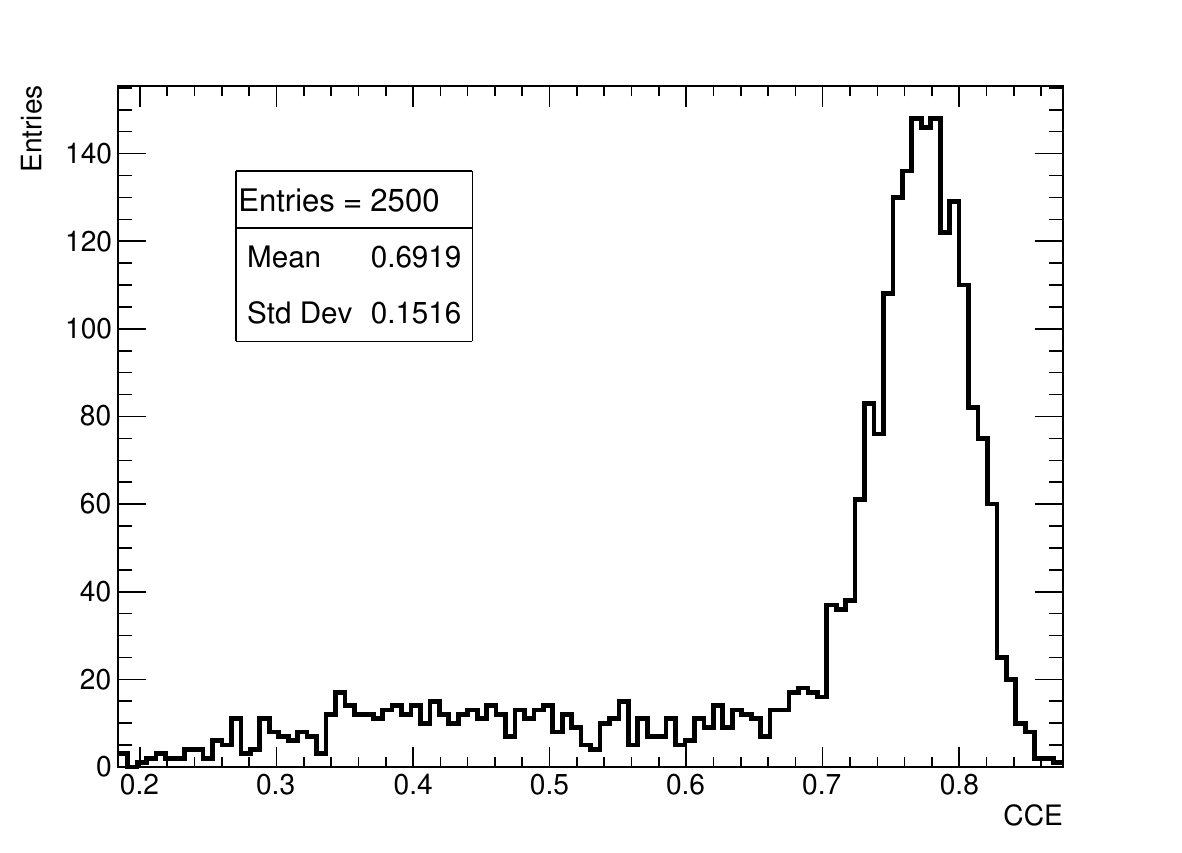} 
\caption{\label{fig:cce_z_1}}
\end{subfigure}
\hfill
\begin{subfigure}{0.49\textwidth}
\centering
\includegraphics[width=0.9\textwidth]{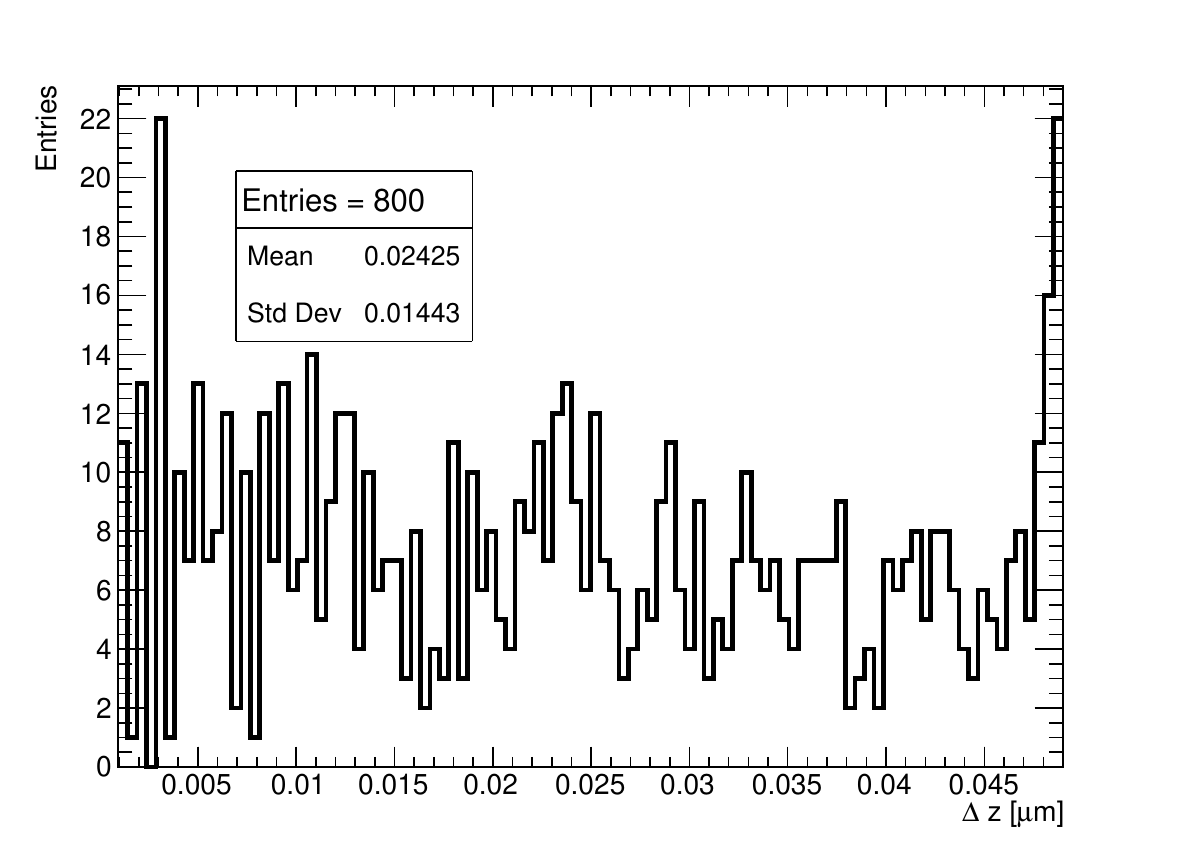}
\caption{\label{fig:delz_z_1}}
\end{subfigure}\\
\hfill
\vfill
\begin{subfigure}{\textwidth}
\centering
\includegraphics[width=0.45\textwidth]{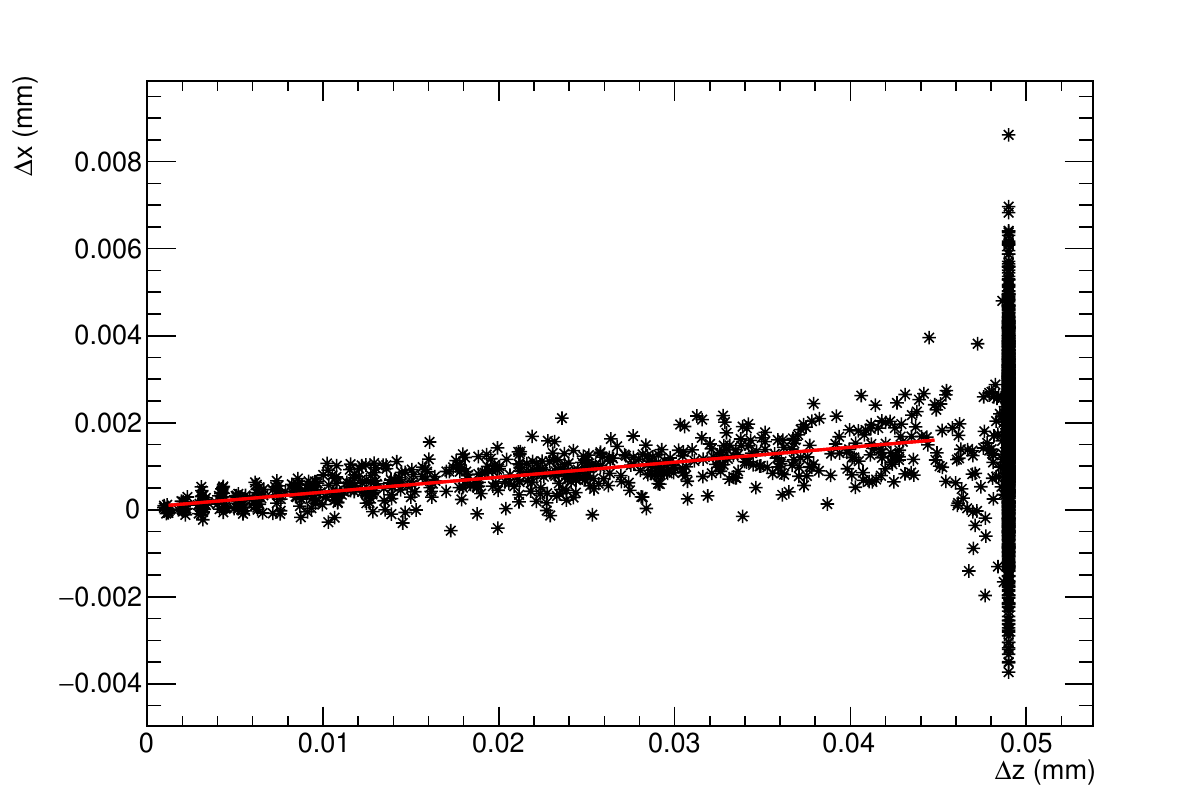} 
\caption{\label{fig:electron_drift}}
\end{subfigure}
\caption{\label{fig:electron_drift_delz}{Distribution} of observables for charge injected close to the sensor midplane. (\textbf{a}) Distribution of simulated CCE. (\textbf{b}) Distribution of simulated $\Delta z$. (\textbf{c}) Visualization of electron drift $\Delta x$ vs. $\Delta z$ towards the pixel side. The red line represents the fit of a straight line to the electron drift, and the average $\tan\theta_{LA}$ is estimated as the slope of the fitted line. }
\end{figure}

 The CCE($z$) is then evaluated by extracting the most probable CCE values (MPV) for each $z$. An example of CCE distribution 
 for charges deposited at  a  $z$ position close to the midplane is reported in 
 Figure \ref{fig:electron_drift_delz}a.

The $\Delta z (z)$ observable is evaluated by taking the average of 
the distribution of the free path traveled by the carriers as a function of the generation depth. This process is illustrated in Figure \ref{fig:electron_drift_delz}b. 

While for the $\Delta z$ distribution, the average over repeated  simulations is a good representative of the distribution itself, it is  not the case for charge collection efficiency; for that, the  most probable value is a better choice.

\begin{figure}[H]
\centering
\begin{subfigure}{0.49\textwidth}
\centering
\includegraphics[width=.9\linewidth]{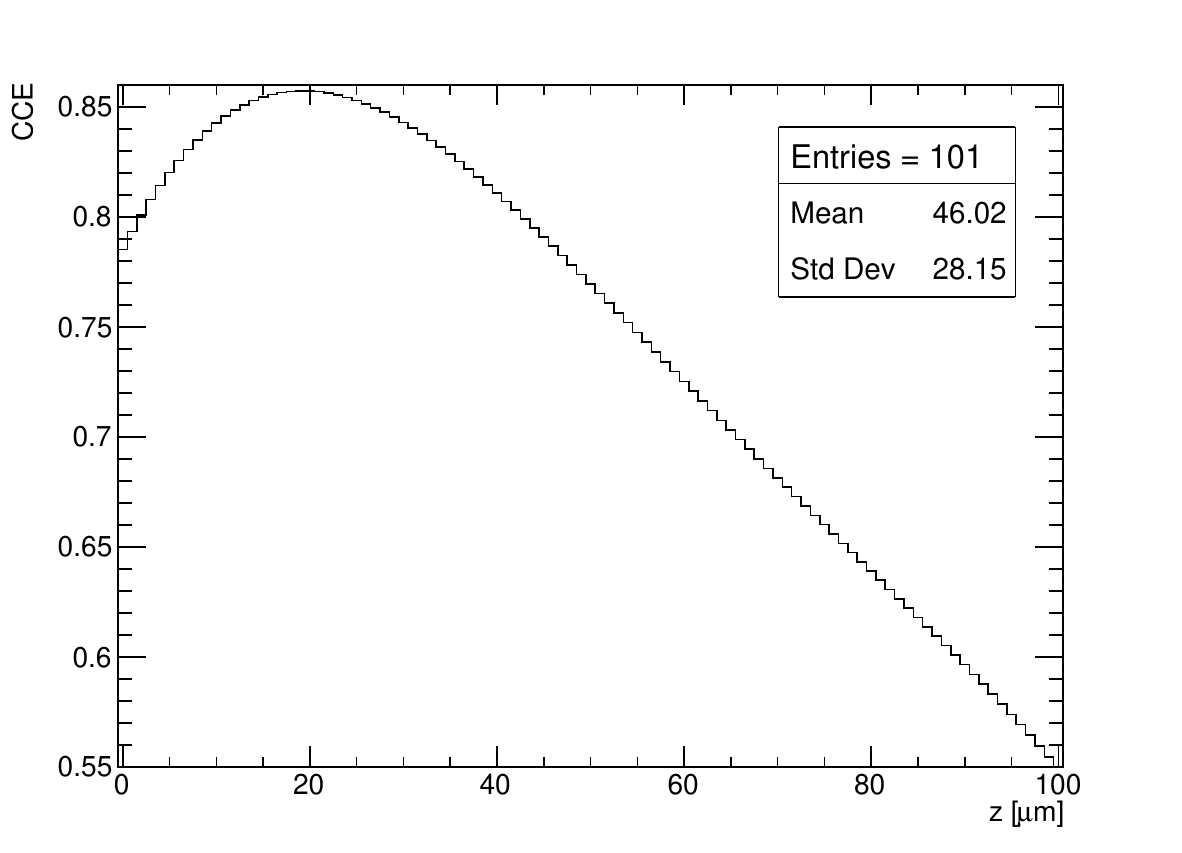}\caption{\label{fig:cce_lut}}
\end{subfigure}
\begin{subfigure}{0.49\textwidth}
\centering
\includegraphics[width=.9\linewidth]{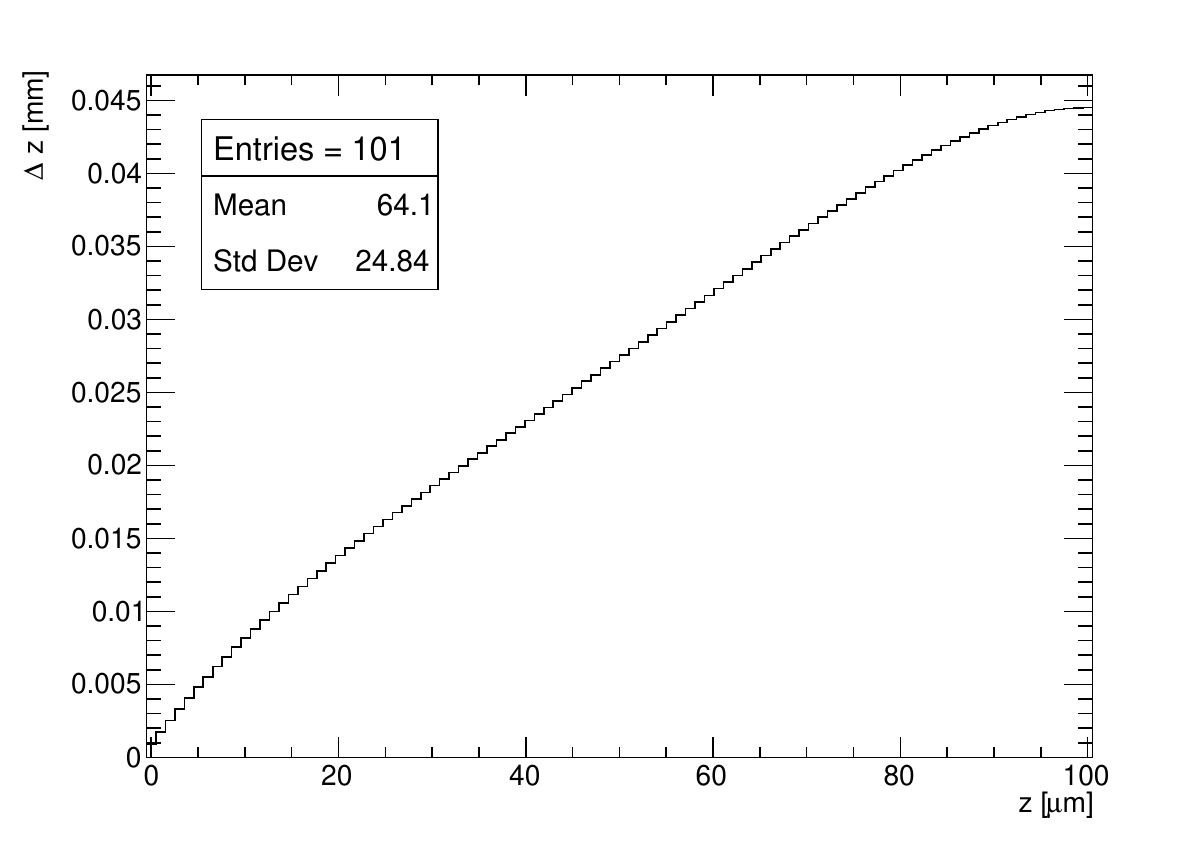}\caption{\label{fig:delz_lut}}
\end{subfigure}\\
\hfill
\vfill
\begin{subfigure}{0.49\textwidth}
\centering
\includegraphics[width=.9\linewidth]{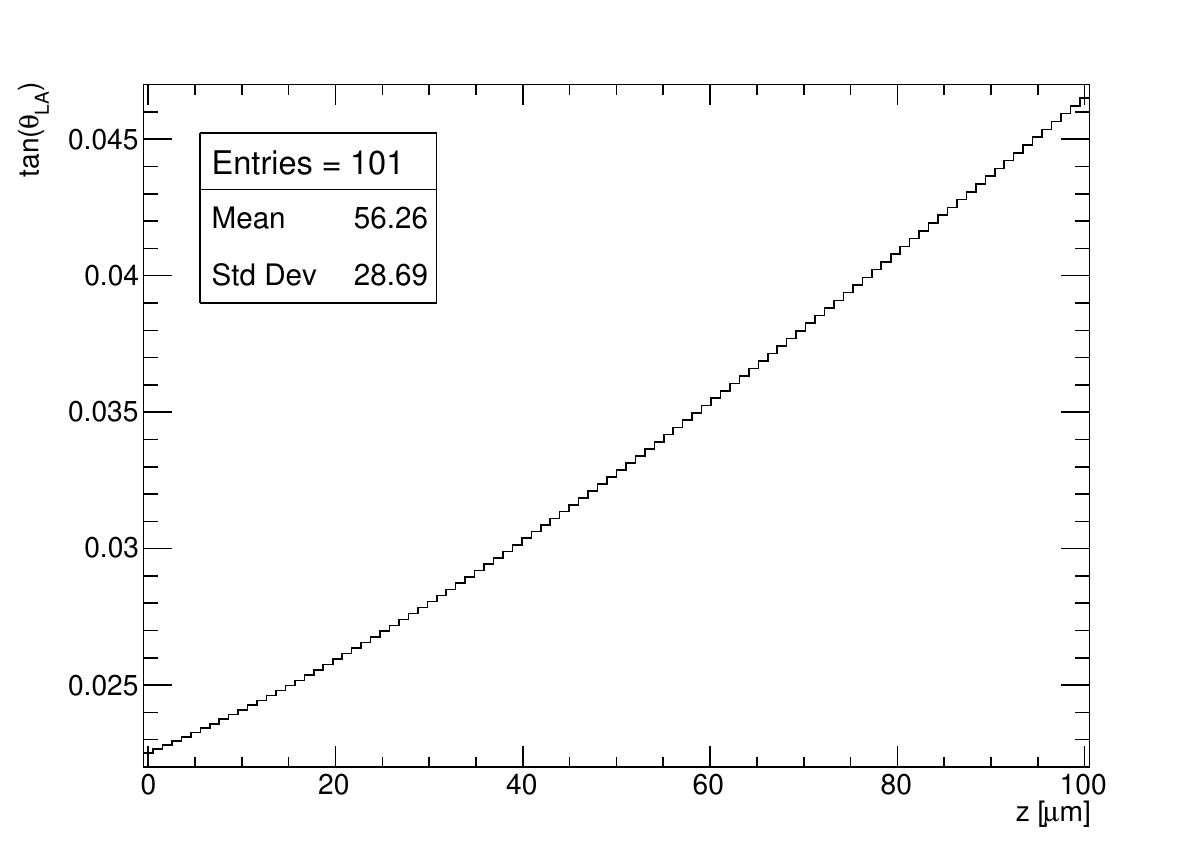}\caption{\label{fig:tanla_lut}}
\end{subfigure}
\caption{LUTs for each of the three observables. The $z$=\SI{0 (100)}{\micro\meter} corresponds to the pixel (backside) position of sensor. (\textbf{a}) CCE($z$); (\textbf{b}) $\Delta z(z)$; (\textbf{c}) $\tan\theta_{LA}(z)$.}
\label{fig:LUTs}
\end{figure}

Finally, the Lorentz angle  $\theta_{LA}(z)$ value is calculated by fitting a straight line to the electron drift distribution $\Delta x$ vs. $\Delta z$ for each $z_{dep}$ position for various $x_{dep}$ and $y_{dep}$ positions, as shown in Figure \ref{fig:electron_drift_delz}c. The slope of the fit at each $z_{dep}$ is then extracted, and the LUT is constructed by plotting the slope ($<\tan(\theta_{LA})>$) as a function of the generation depth, as illustrated in 
Figure \ref{fig:LUTs}c.

The large spread of $\Delta x$ values observed when electrons reach the electrode is an artifact of the simulation. At this position, electrons are already within the pixel implant and cease to diffuse further. To prevent biasing the results, the upper limit of the fit range for extracting $\theta_{LA}(z)$ is constrained to a few micrometers away from the pixel implant.

Finally, Figure \ref{fig:LUTs} displays the  LUTs for charge collection efficiency (CCE(z)),   average free path ($\Delta z (z)$), and tangent of Lorentz angle ($\tan\theta_{LA}(z)$) 
for the case presented in this paper.

The CCE peaks at over 85\% at approximately \SI{20}{\micro\meter} below the pixel implant. This phenomenon occurs because holes become trapped near the pixel, thereby partially screening a portion of the signal. However, the influence of holes becomes negligible when charge is deposited farther away from the pixel electrode.

The average distance $\Delta z$ covered by electrons becomes  significantly less than the distance to the electrode rather  quickly. For instance, at $z = $ \SI{20}{\micro\meter}, the average distance $\Delta z$ is approximately  \SI{15}{\micro\meter}, which is 25~\% less.

The Lorentz angle deflection is predicted to be very small everywhere  in the bulk compared to the typical value in the ATLAS sensors  before irradiation---about 0.22 radians. The main reason for this  is the large electric field inside the sensor, especially at the  pixel side. This reduction in the Lorentz angle after irradiation  has already been observed with collision data in ATLAS pixels \cite{LlorenteMerino:2655892}. The largest deflection in $x$ will be for charges  deposited at the backside, and it will be just $ \SI{2}{\micro\meter}$. 

The following section will present the first validation of the LUT algorithm.


\section{Internal Validation of the LUTs Digitizer: Closure Test}
\label{sec:closure}

A first validation of the new algorithm has been carried out through a closure test.  As explained above in Allpix$^2$, it is possible to implement 
different algorithms for each of the various steps of the  simulation. In particular, after the energy deposition step (see Figure \ref{fig:APSQ_SimChain_RadDam}), the charge transport 
can be modeled in several ways, depending on the needed accuracy. 

This feature was exploited for the purpose of our closure test  by implementing a new charge transport algorithm  based on LUTs, the ``LUT propagator''. This propagator is based on Equation (\ref{eq:rprop}). For every deposited electron the induced signal and final position is evaluated using the LUTs. 

In the closure test presented, events simulated using the LUT propagator have been compared to fully simulated (FS) events.

Events were simulated using sensors of the same type and under conditions used to generate the LUTs, as discussed in Section \ref{sec:lut}. The agreement between LUT-based and full simulation (FS) events are assessed by examining cluster charge and cluster size, independently in the transverse ($x$) and longitudinal ($y$) directions {(ATLAS uses a right-handed coordinate system with its origin at the nominal interaction point (IP)} in the center of the detector and the z-axis coinciding with the axis of the beam pipe. The x-axis points from the IP towards the center of the LHC ring, and the y-axis points upward. Cylindrical coordinates (r,$\phi$) are used in the transverse plane, $\phi$ being the azimuthal angle around the beam axis. The pseudorapidity is defined in terms of the polar angle $\theta$ as $\eta = - \ln \tan(\theta/2)$).

For charge deposition, a beam of pions at three distinct energies is simulated, corresponding to transverse momentum ($p_T$) values of 1, 10, and 100 GeV/c.

In order to reproduce the configuration in which the simulated pixel  sensors will be installed, the module is tilted around the 
beam axis by 0.25 radians and at different $\eta$ values, from 0 
(normal incidence) to 1.4 (about 1.1 radian with respect to the 
normal). 

\textls[-20]{In the following, the results for  cluster charge (Section \ref{sec:closure_cluster_charge}) and for 
cluster size  (Section \ref{sec:closure_cluster_size})  for $p_T$ = 1 GeV/c  will be presented. Results will be summarized 
for all $p_T$ values in Section \ref{sec:closure_summary}, 
accompanied by a brief discussion.}

\subsection{Results of the Closure Test on Cluster Charge}
\label{sec:closure_cluster_charge}

\begin{figure}[H]
\centering
\begin{subfigure}{0.49\textwidth}
\includegraphics[width=0.95\linewidth]{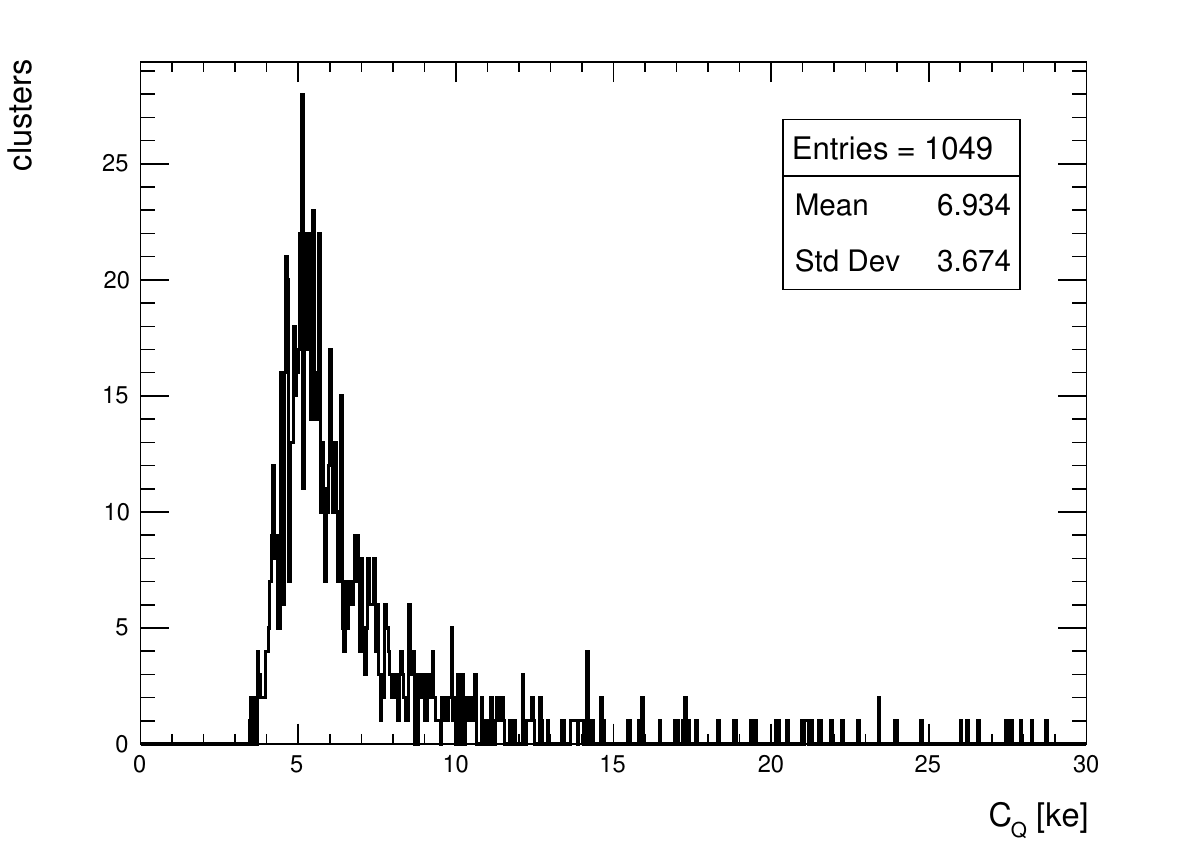} 
\caption{\label{fig:closure_Charge_1GeV_0Eta_FS}}

\end{subfigure}
\begin{subfigure}{0.49\textwidth}
\includegraphics[width=0.95\linewidth]{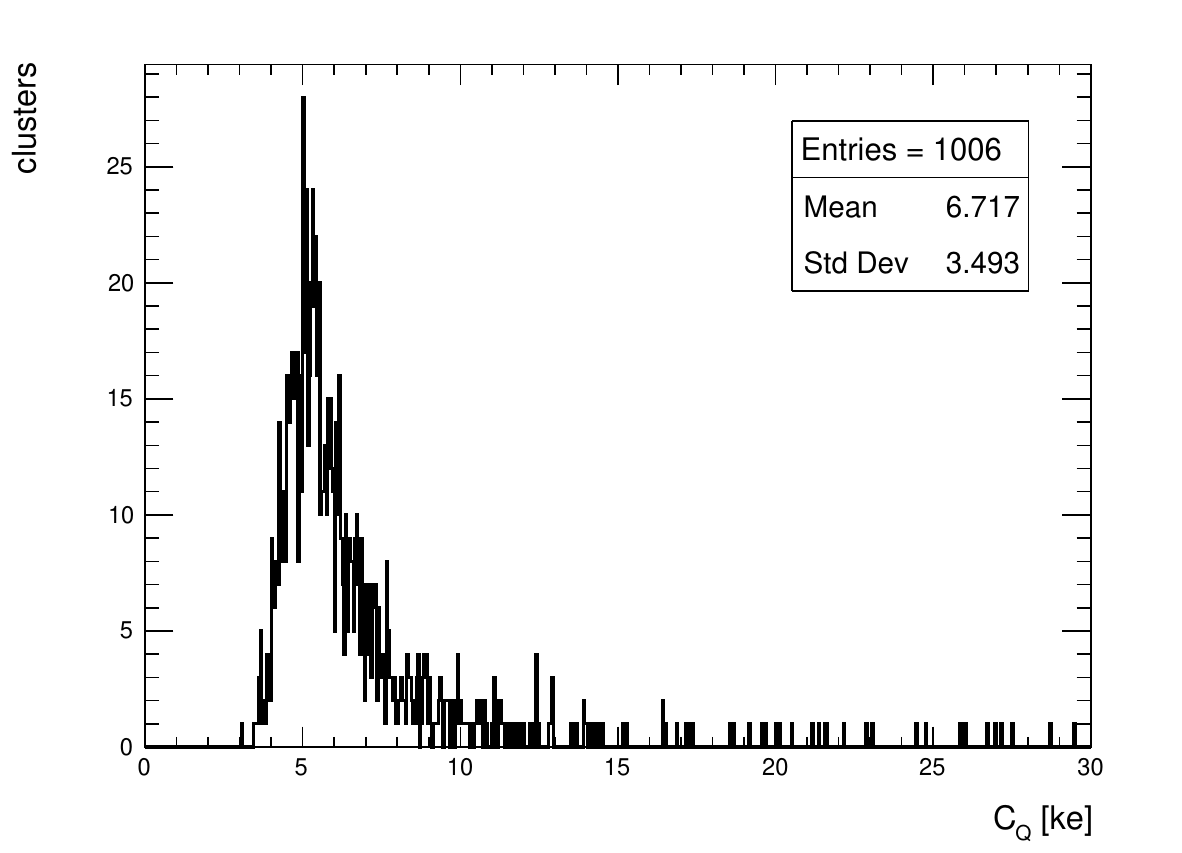}
\caption{\label{fig:closure_Charge_1GeV_0Eta_LUT}}
\end{subfigure}
\caption{\label{fig:closure_Charge_1GeV_0Eta}Cluster charge 
distribution for pions with $p_T$ = 1 GeV/c impinging at $\eta$ = 0. (\textbf{a}) Full simulation; (\textbf{b}) LUT-based simulation.}
\end{figure}

\begin{figure}[H]
\centering
\begin{subfigure}{0.49\textwidth}
\includegraphics[width=0.95\linewidth]{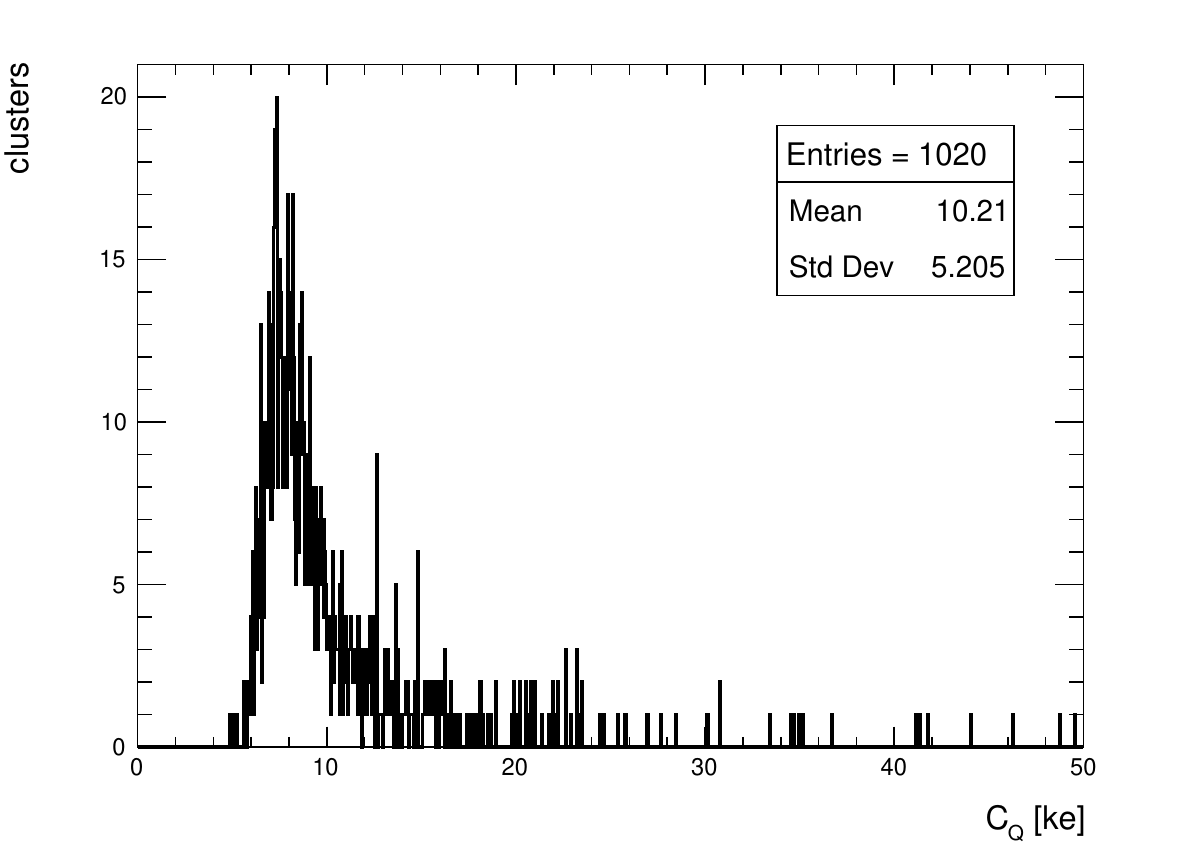} 
\caption{\label{fig:closure_Charge_1GeV_1Eta_FS}}

\end{subfigure}
\begin{subfigure}{0.49\textwidth}
\includegraphics[width=0.95\linewidth]{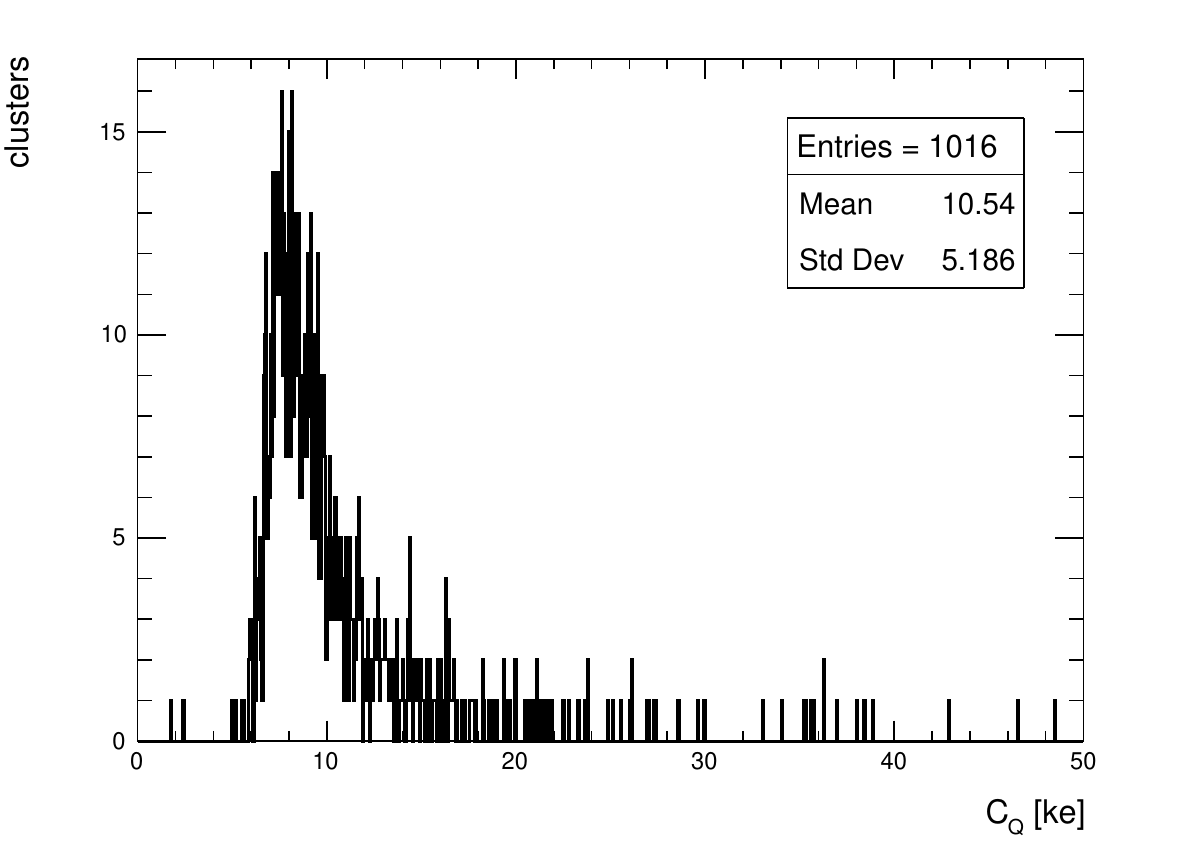}
\caption{\label{fig:closure_Charge_1GeV_1Eta_LUT}}
\end{subfigure}
\caption{\label{fig:closure_Charge_1GeV_1Eta}Cluster charge 
distribution for pions with $p_T$ = 1 GeV/c impinging at $\eta$ = 1. (\textbf{a}) Full simulation; (\textbf{b}) LUT-based simulation.}
\end{figure}

\begin{figure}[H]
\centering
\begin{subfigure}{0.49\textwidth}
\includegraphics[width=0.95\linewidth]{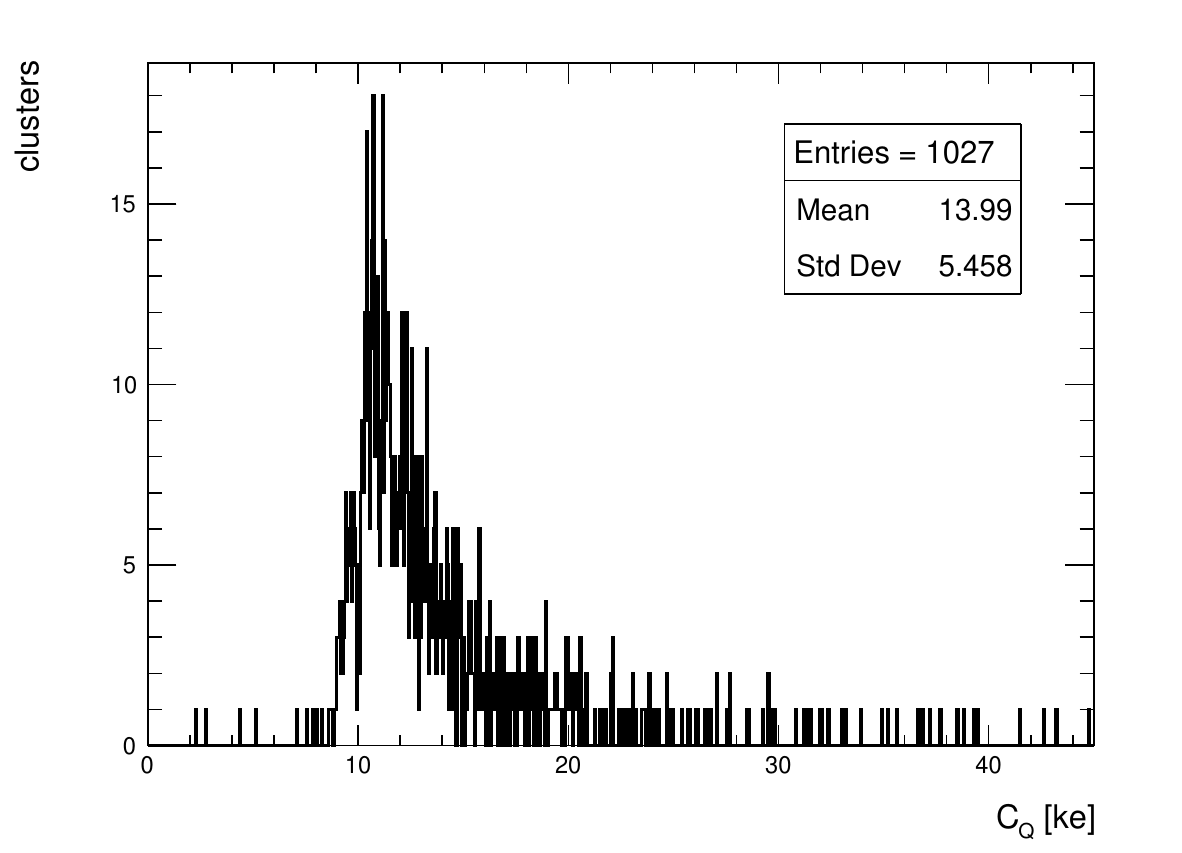} 
\caption{\label{fig:closure_Charge_1GeV_1_4Eta_FS}}

\end{subfigure}
\begin{subfigure}{0.49\textwidth}
\includegraphics[width=0.95\linewidth]{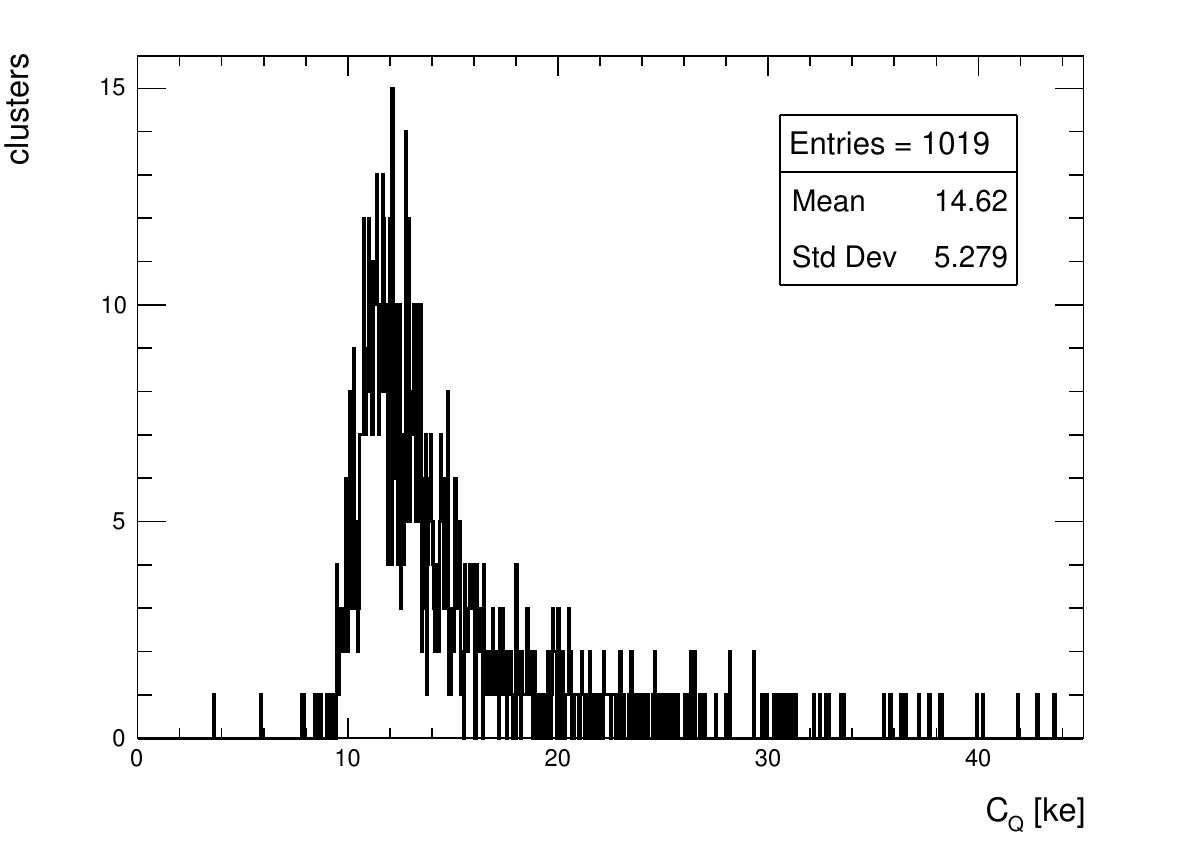}
\caption{\label{fig:closure_Charge_1GeV_1_4Eta_LUT}}
\end{subfigure}
\caption{\label{fig:closure_Charge_1GeV_1_4Eta}Cluster charge 
distribution for pions with $p_T$ = 1 GeV/c impinging at $\eta$ = 1.4. (\textbf{a}) Full simulation; (\textbf{b}) LUT-based simulation.}
\end{figure}

The cluster charge distributions from simulated events for  $p_T$ = 1 GeV/c pions are shown in Figures \ref{fig:closure_Charge_1GeV_0Eta}--\ref{fig:closure_Charge_1GeV_1_4Eta} for $\eta$ = 0, 1 and 
1.4, respectively, for both  the full simulation and the LUT-based simulation. 

The level of agreement between FS and LUT simulation is remarkable:  the agreement between the mean values of the two distributions is
 well within 5\%. Similar results are obtained for $p_T$ = 10 and 
 100 GeV/c.
 
\unskip

\subsection{Results of the Closure Test on Cluster Size}
\label{sec:closure_cluster_size}

\begin{figure}[H]
\centering
\begin{subfigure}{0.49\textwidth}
\includegraphics[width=0.95\linewidth]{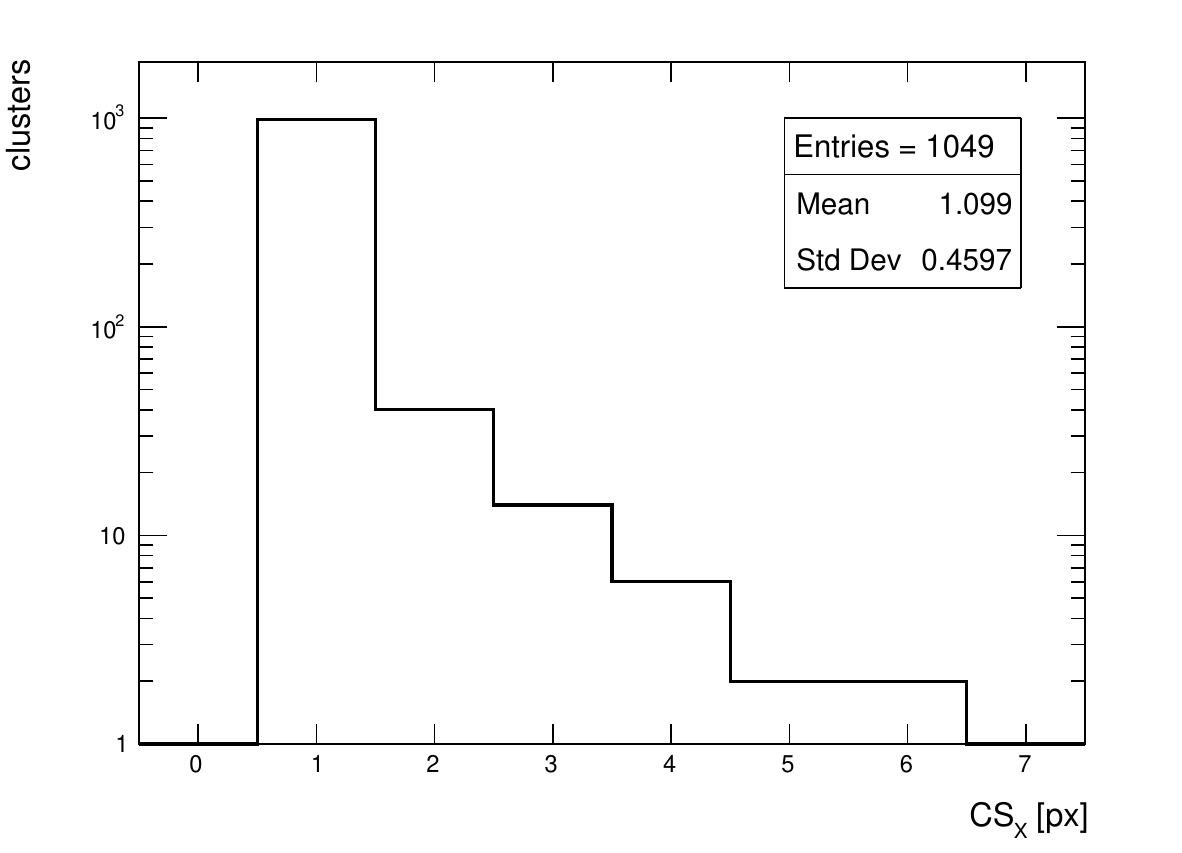} 
\caption{\label{fig:closure_CSX_1GeV_0Eta_FS}}

\end{subfigure}
\begin{subfigure}{0.49\textwidth}
\includegraphics[width=0.95\linewidth]{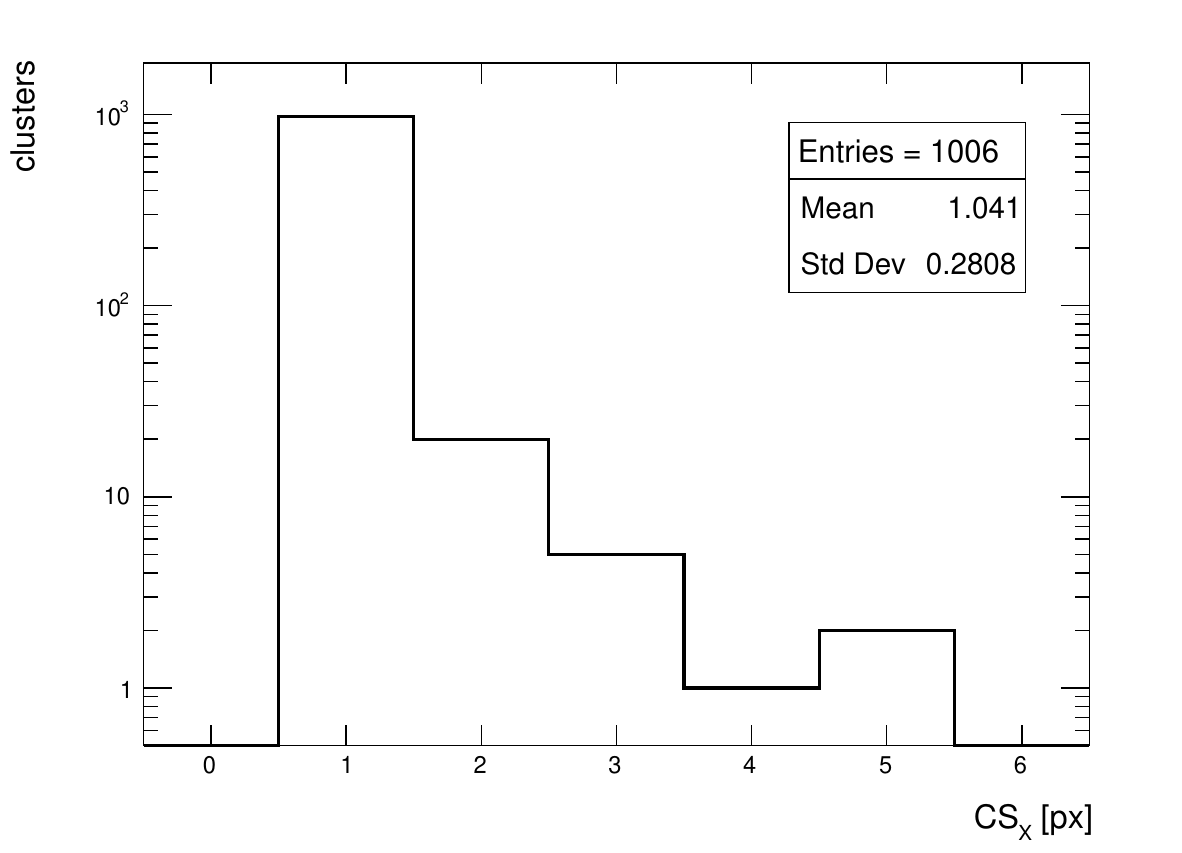}
\caption{\label{fig:closure_CSX_1GeV_0Eta_LUT}}
\end{subfigure}
\caption{\label{fig:closure_CSX_1GeV_0Eta}Cluster size  
distribution in the transverse direction for pions with $p_T$ = 1 GeV/c impinging at $\eta$ = 0. (\textbf{a}) Full simulation; (\textbf{b}) LUT-based simulation.}
\end{figure}

\begin{figure}[H]
\centering
\begin{subfigure}{0.49\textwidth}
\includegraphics[width=0.95\linewidth]{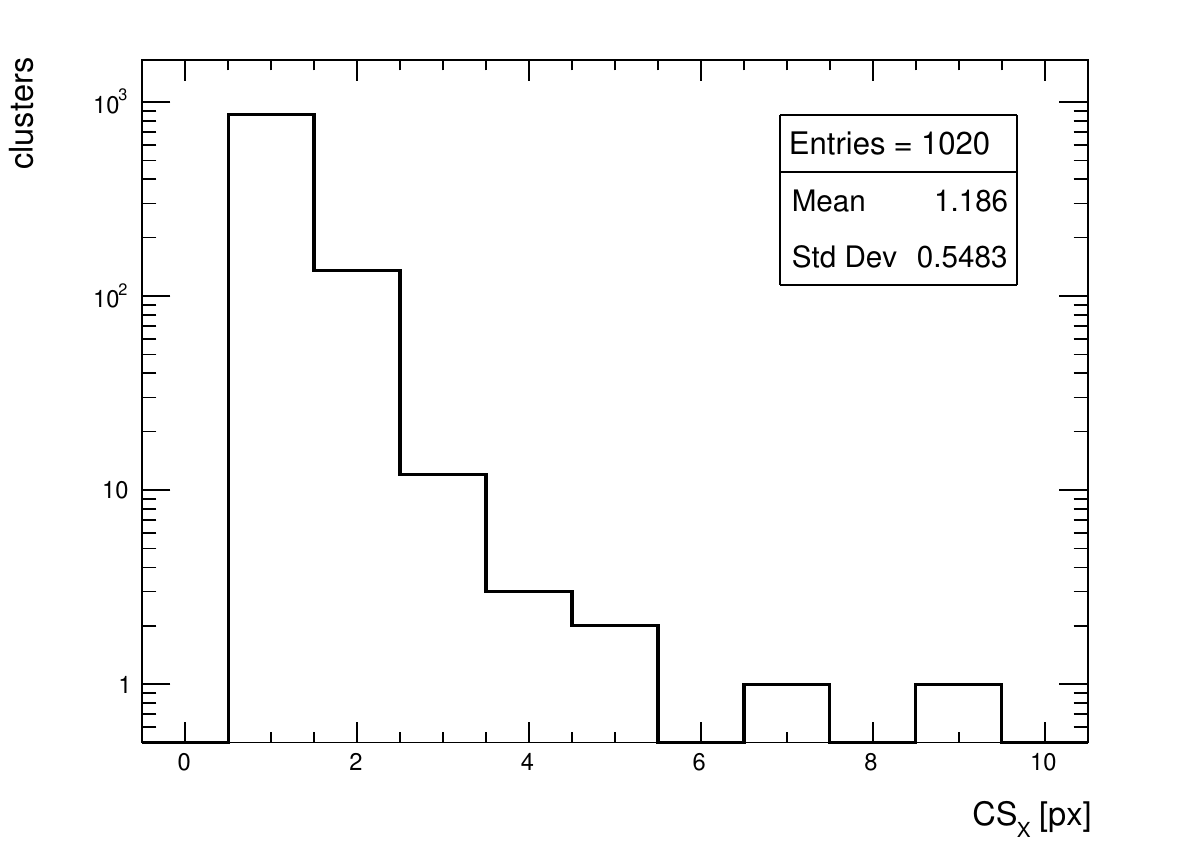} 
\caption{\label{fig:closure_CSX_1GeV_1Eta_FS}}

\end{subfigure}
\begin{subfigure}{0.49\textwidth}
\includegraphics[width=0.95\linewidth]{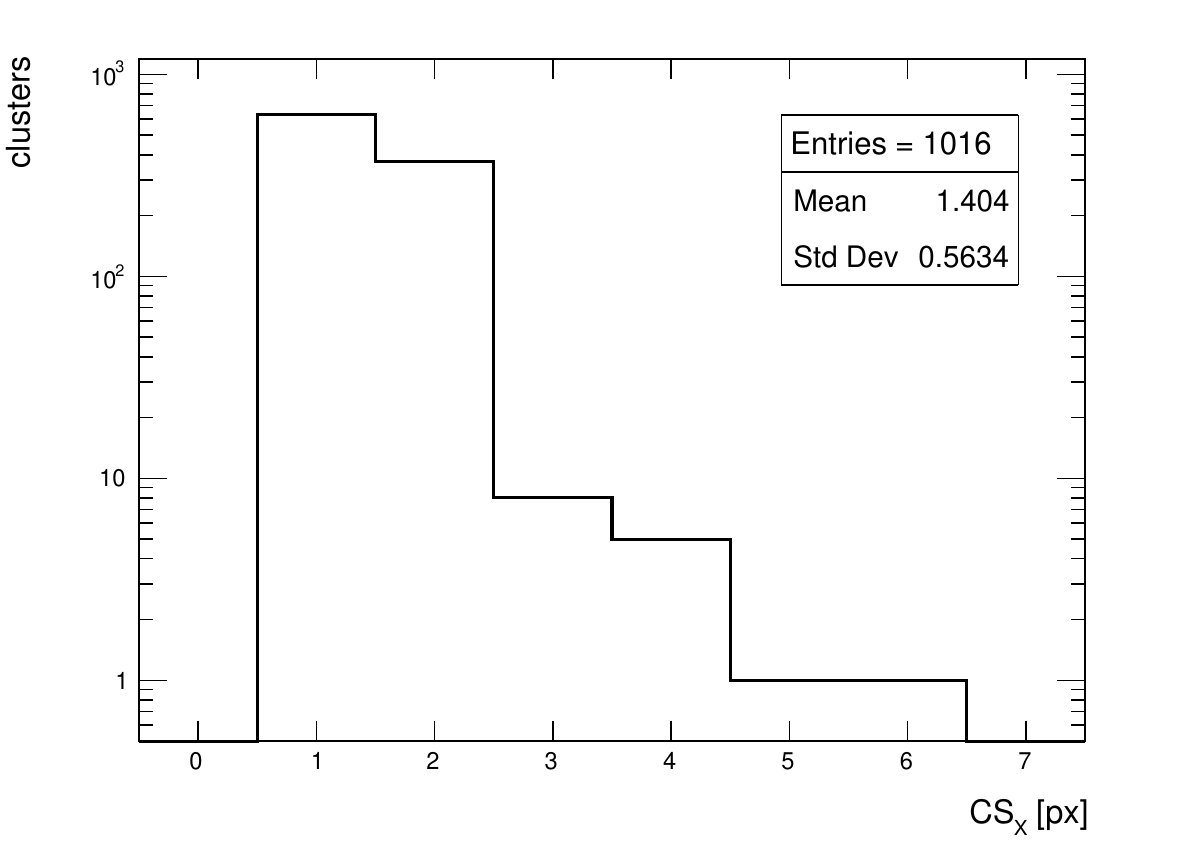}
\caption{\label{fig:closure_CSX_1GeV_1Eta_LUT}}
\end{subfigure}
\caption{\label{fig:closure_CSX_1GeV_1Eta}Cluster size  
distribution in the transverse direction for pions with $p_T$ = 1 GeV/c impinging at $\eta$ = 1. (\textbf{a}) Full simulation; (\textbf{b}) LUT-based simulation.}
\end{figure}

\begin{figure}[H]
\centering
\begin{subfigure}{0.49\textwidth}
\includegraphics[width=0.95\linewidth]{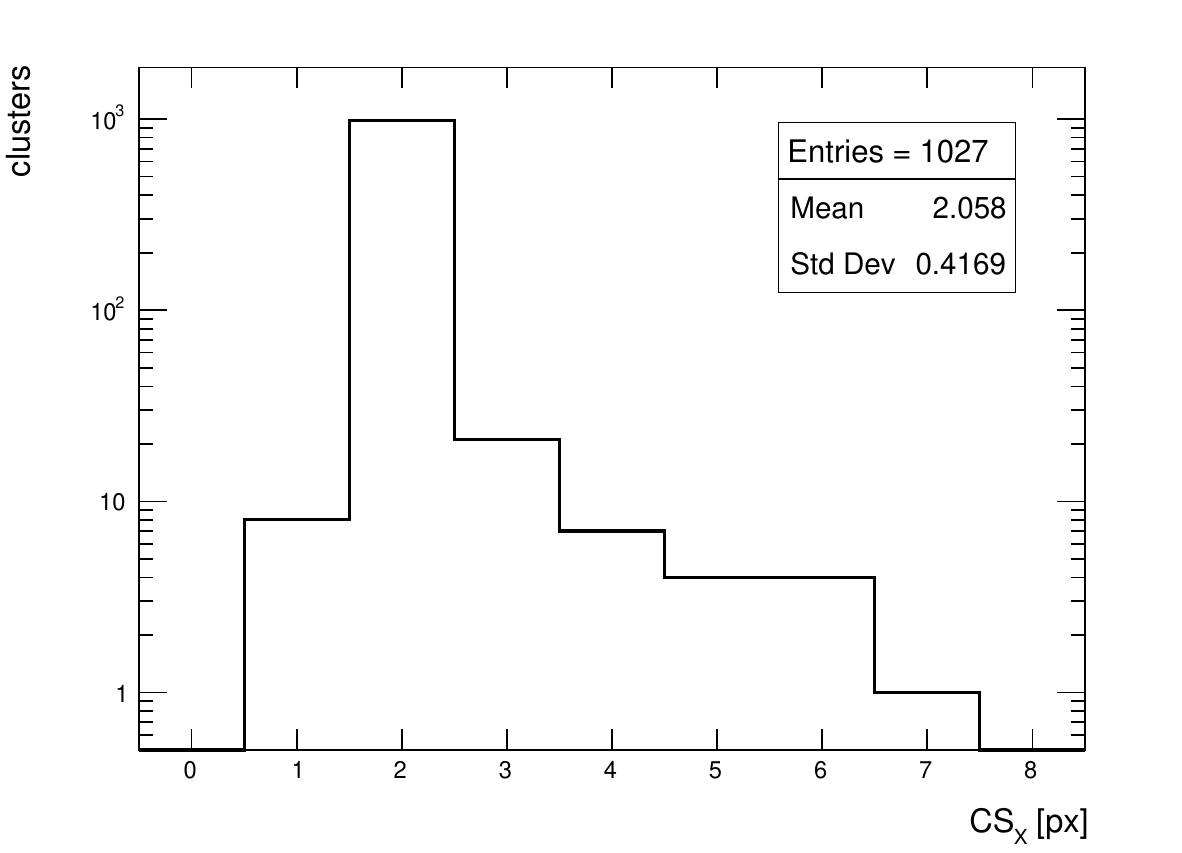} 
\caption{\label{fig:closure_CSX_1GeV_1_4Eta_FS}}

\end{subfigure}
\begin{subfigure}{0.49\textwidth}
\includegraphics[width=0.95\linewidth]{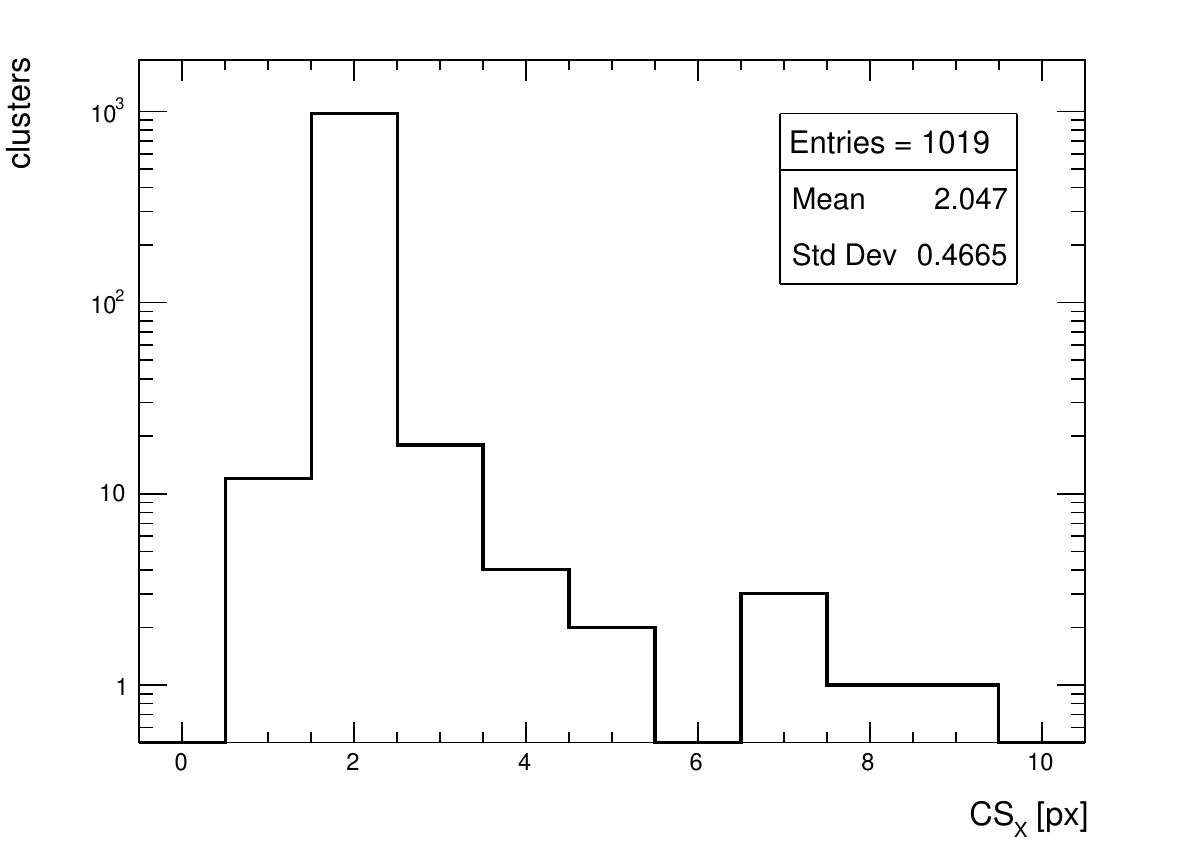}
\caption{\label{fig:closure_CSX_1GeV_1_4Eta_LUT}}
\end{subfigure}
\caption{\label{fig:closure_CSX_1GeV_1_4Eta}Cluster size  
distribution in the transverse direction for pions with $p_T$ = 1 GeV/c impinging at $\eta$ = 1.4. (\textbf{a}) Full simulation; (\textbf{b}) LUT-based simulation.}
\end{figure}

The cluster size distributions in the transverse direction from simulated events for  \linebreak $p_T$ = 1 GeV/c pions are shown in Figures \ref{fig:closure_CSX_1GeV_0Eta}--\ref{fig:closure_CSX_1GeV_1_4Eta} for $\eta$ = 0, 1 and 
1.4, respectively, for both  full simulation and LUT-based simulation. 

\unskip
\unskip

\begin{figure}[H]
\centering
\begin{subfigure}{0.49\textwidth}
\includegraphics[width=0.9\linewidth]{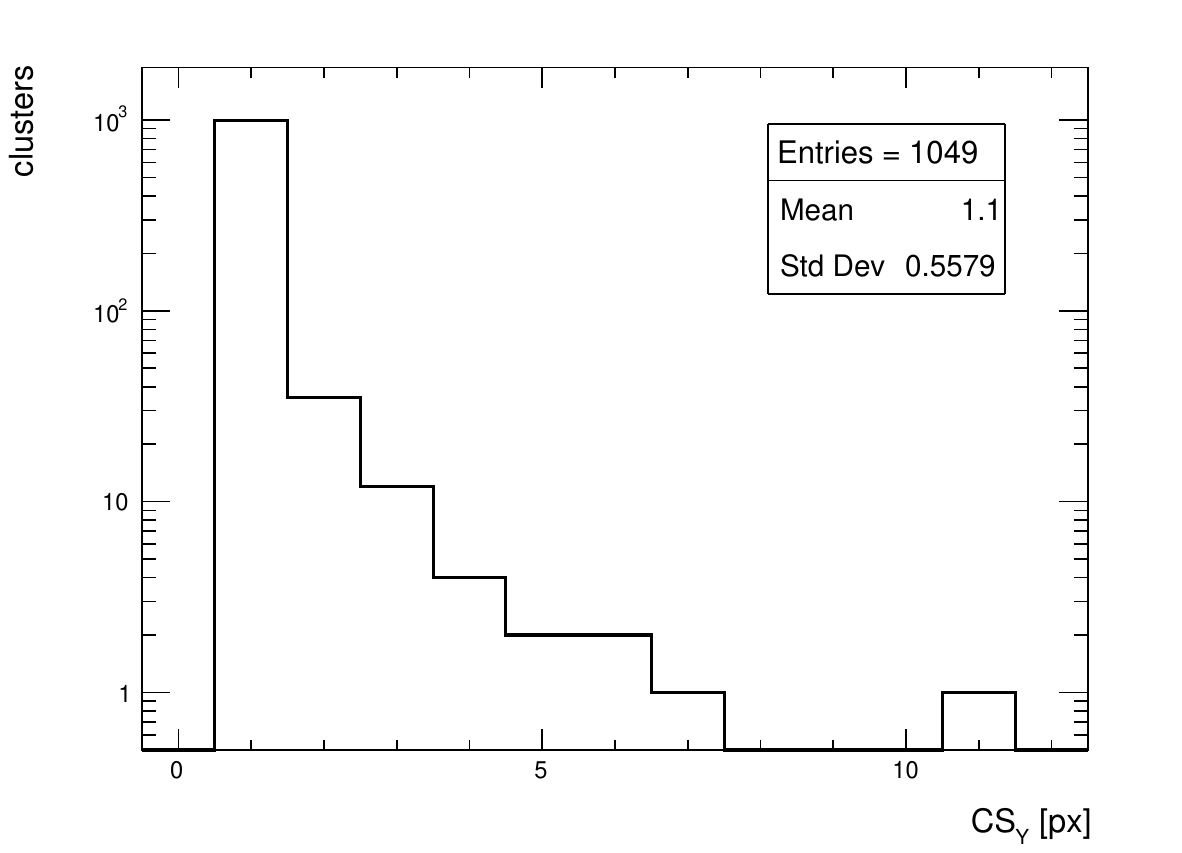} 
\caption{\label{fig:closure_CSY_1GeV_0Eta_FS}}

\end{subfigure}
\begin{subfigure}{0.49\textwidth}
\includegraphics[width=0.9\linewidth]{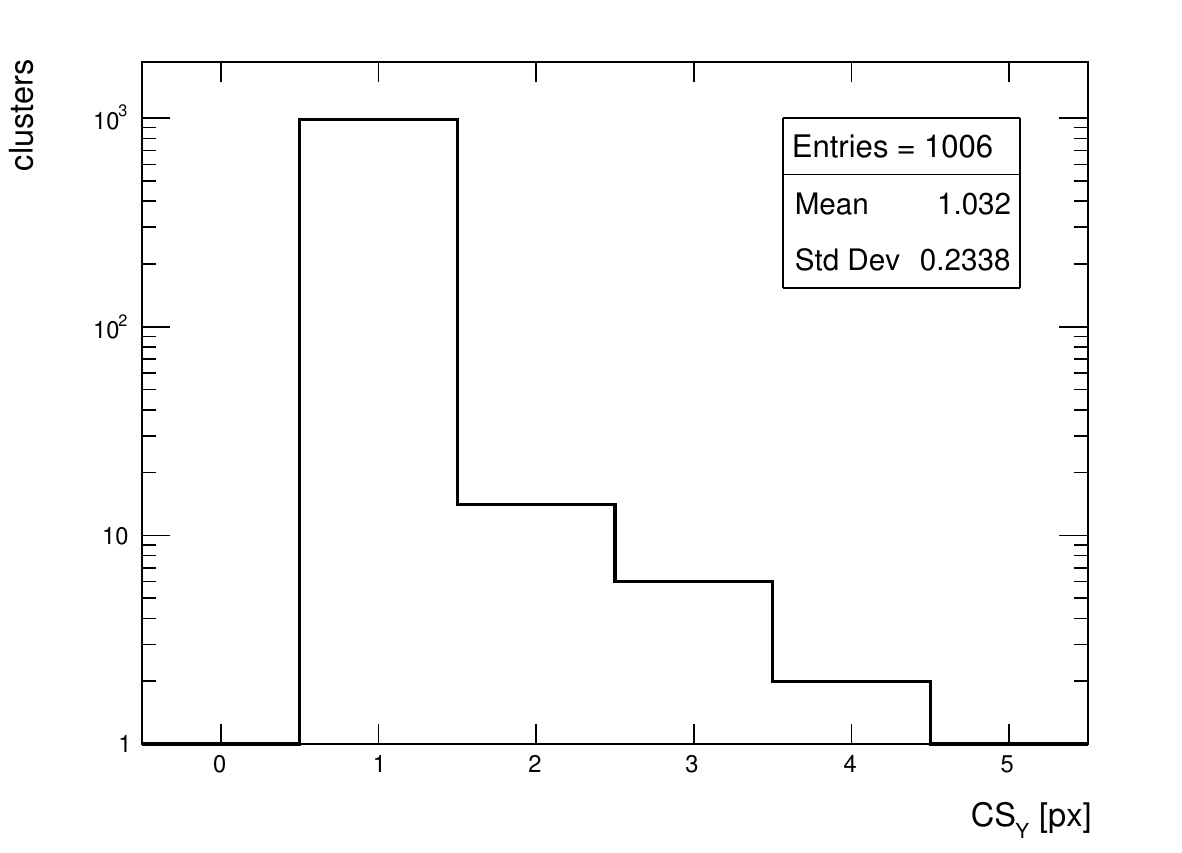}
\caption{\label{fig:closure_CSY_1GeV_0Eta_LUT}}
\end{subfigure}
\caption{\label{fig:closure_CSY_1GeV_0Eta}Cluster size  
distribution in the longitudinal direction for pions with $p_T$ = 1 GeV/c impinging at $\eta$ = 0. (\textbf{a}) Full simulation; (\textbf{b}) LUT-based simulation.}
\end{figure}

\begin{figure}[H]
\centering
\begin{subfigure}{0.49\textwidth}
\includegraphics[width=0.9\linewidth]{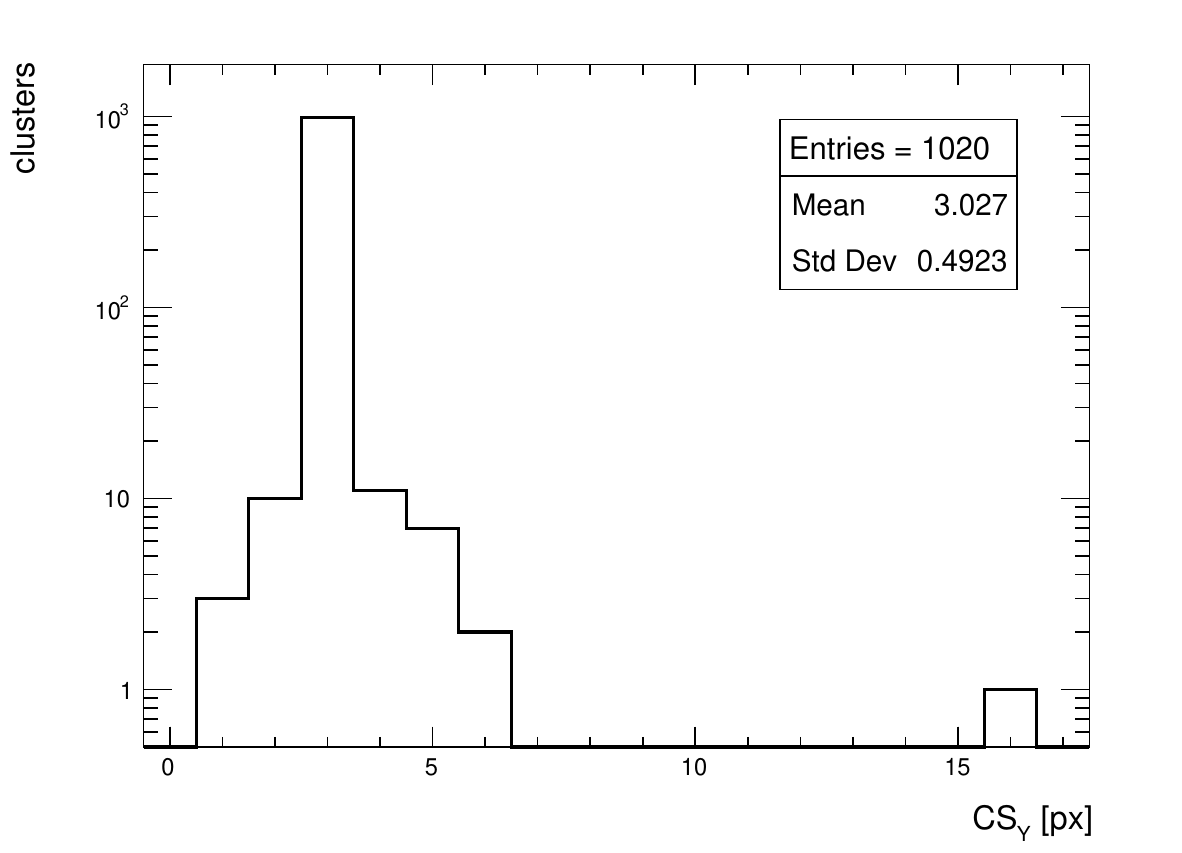} 
\caption{\label{fig:closure_CSY_1GeV_1Eta_FS}}

\end{subfigure}
\begin{subfigure}{0.49\textwidth}
\includegraphics[width=0.9\linewidth]{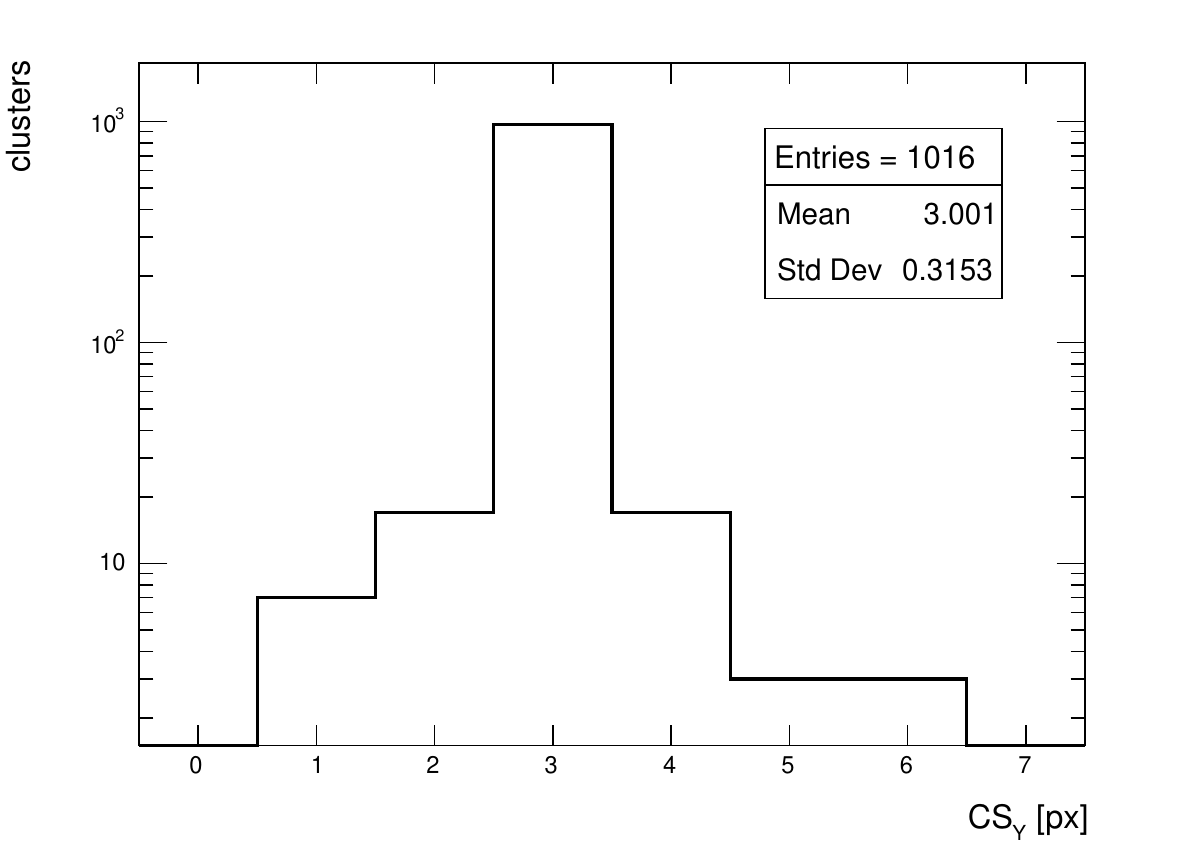}
\caption{\label{fig:closure_CSY_1GeV_1Eta_LUT}}
\end{subfigure}
\caption{\label{fig:closure_CSY_1GeV_1Eta}Cluster size  
distribution in the longitudinal direction for pions with $p_T$ = 1 GeV/c impinging at $\eta$ = 1. (\textbf{a}) Full simulation; (\textbf{b}) LUT-based simulation.}
\end{figure}

\begin{figure}[H]
\centering
\begin{subfigure}{0.49\textwidth}
\includegraphics[width=0.9\linewidth]{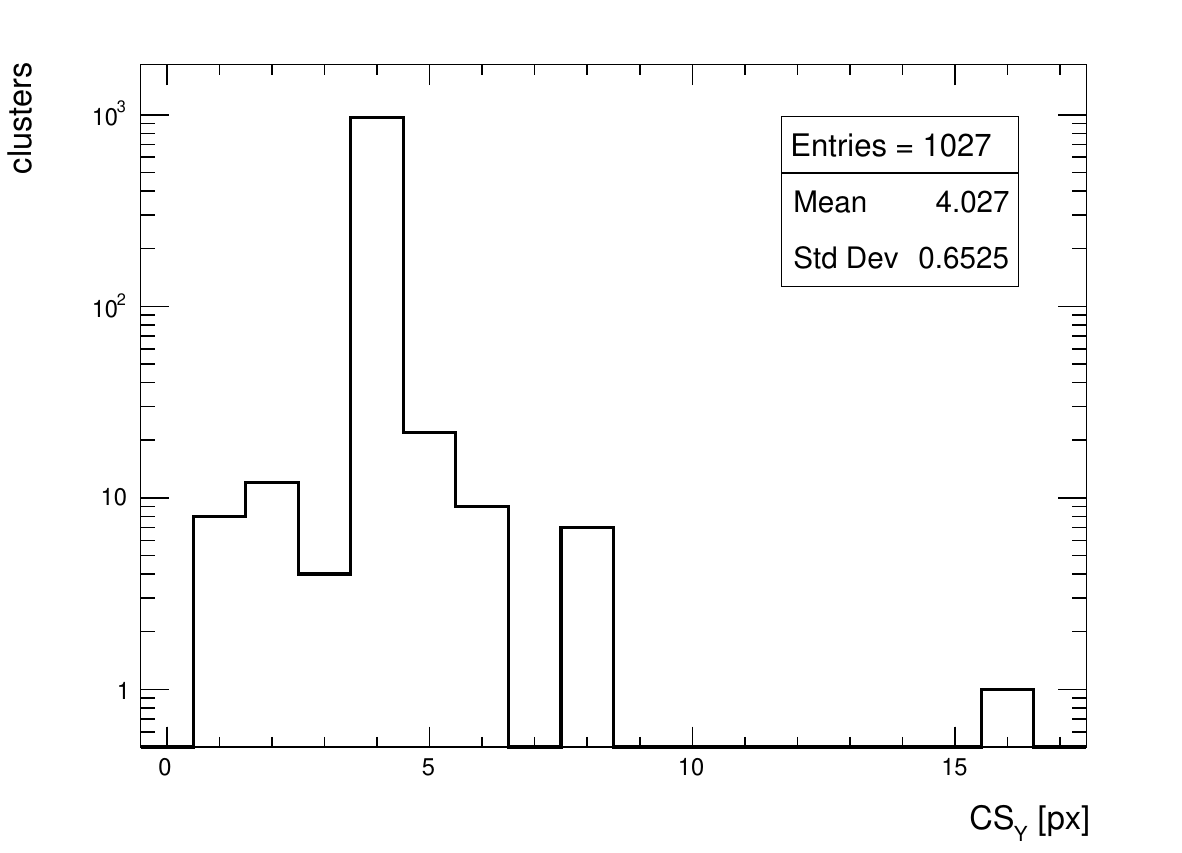} 
\caption{\label{fig:closure_CSY_1GeV_1_4Eta_FS}}

\end{subfigure}
\begin{subfigure}{0.49\textwidth}
\includegraphics[width=0.9\linewidth]{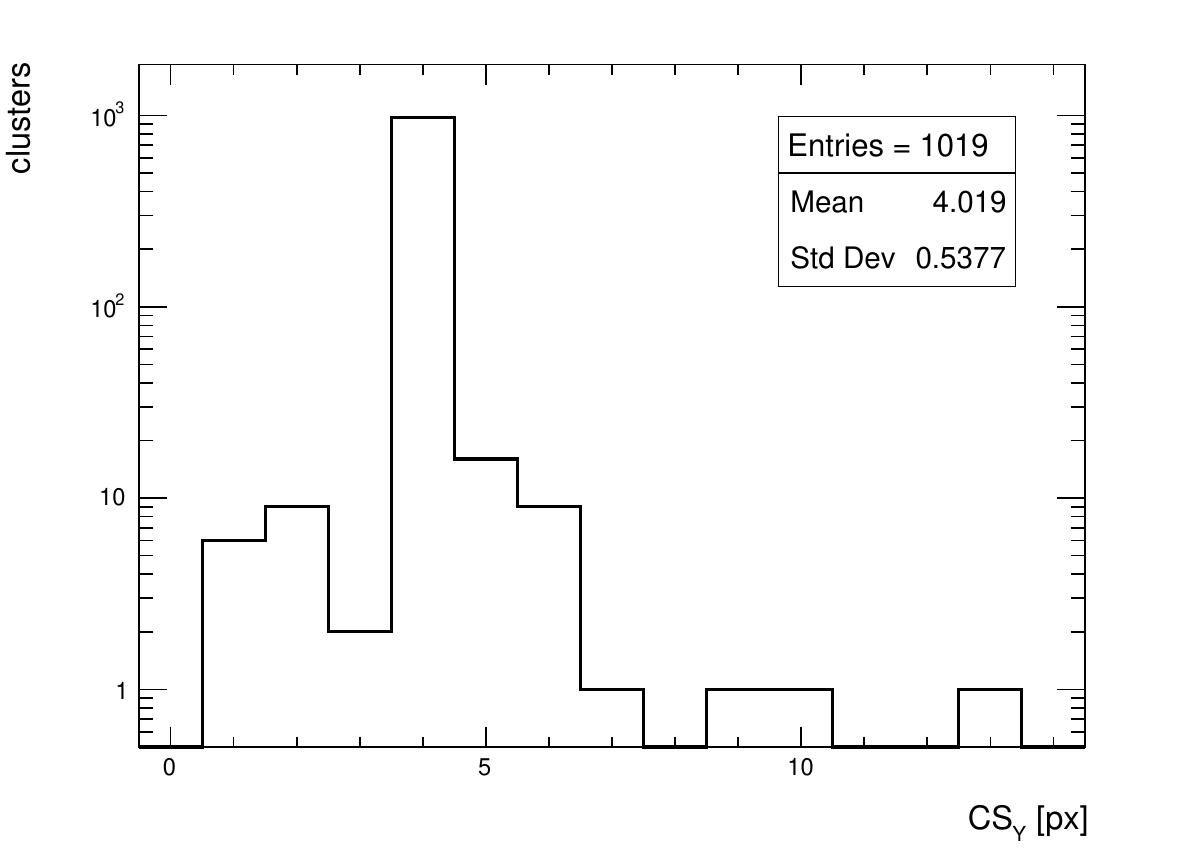}
\caption{\label{fig:closure_CSY_1GeV_1_4Eta_LUT}}
\end{subfigure}
\caption{\label{fig:closure_CSY_1GeV_1_4Eta}Cluster size  
distribution in the longitudinal direction for pions with $p_T$ = 1 GeV/c impinging at $\eta$ = 1.4. (\textbf{a}) Full simulation; (\textbf{b}) LUT-based simulation.}
\end{figure}

The cluster size distributions in the longitudinal direction from simulated events for  $p_T$ = 1 GeV/c pions are shown in Figures \ref{fig:closure_CSY_1GeV_0Eta}--\ref{fig:closure_CSY_1GeV_1_4Eta} for $\eta$ = 0, 1 and 
1.4, respectively, for both  full simulation and LUT-based simulation. 

Despite an overall good agreement between the average of the 
cluster size distribution, some tension exists in the transverse 
distribution for $\eta = 1$. The difference is of the order of 18\%, while for the other $\eta$ 
values the agreement is within few \%, as seen for  the cluster charge. Similar behavior is observed for $p_T$ = 10 and 100 GeV/c.

\unskip

\subsection{Summary and Discussion on Closure Test}
\label{sec:closure_summary}

In Tables \ref{tab:closure_1GeV}--\ref{tab:closure_100GeV}, the summary of the results of 
the closure tests is presented for all quantities and pseudorapidities for pions with $p_T$ = 1, 10 and 100 GeV/c, respectively.


\begin{table}[H]
\caption{\label{tab:closure_1GeV}Average cluster size in transverse 
($CS_X$) and bending ($CS_Y$) direction, and average cluster charge $C_Q$ for FS- and LUT-based events for different pseudorapidity values $\eta$ and $p_T$ = 1 GeV/c; $\epsilon$ is the relative difference between FS and LUT.}
\newcolumntype{C}{>{\centering\arraybackslash}X}
\begin{tabularx}{\textwidth}{CCCccccccc}
\toprule
 \multirow[b]{1.78}{*}{\boldmath{$\eta$}} & \multicolumn{3}{c}{\boldmath{$CS_X$}} & \multicolumn{3}{c}{\boldmath{$CS_Y$}} & \multicolumn{3}{c}{\boldmath{$C_Q$\textbf{ [ke]}}} \\
   
   \cmidrule{2-10}
      & \textbf{FS}& \textbf{LUT} & \textbf{$\epsilon$ [\%]}& \textbf{FS} & \textbf{LUT} & \textbf{\textbf{$\epsilon$} [\%]}&\textbf{FS}& \textbf{LUT} & \textbf{$\epsilon$ [\%]} \\
      \midrule
      0 & 1.099 & 	1.041 &	5.3 & 1.1	& 1.032 &	6.2		& 6.957	& 6.741	& 3.1 \\
      \midrule
     1  &	1.186	& 1.404	& 18.38	& 3.027 &	3.001	& 0.86 & 10.25	& 10.54 &	2.8\\
     \midrule
     1.4 & 2.058	& 2.047	& 0.53 & 4.027	& 4.019	& 0.20 & 
     13.99 &	14.62	& 4.5 \\
     \bottomrule
\end{tabularx}
\end{table}


\begin{table}[H]
\caption{\label{tab:closure_10GeV}Average cluster size in transverse 
($CS_X$) and bending ($CS_Y$) direction, and average cluster charge $C_Q$ for FS- and LUT-based events for different pseudorapidity values $\eta$ and $p_T$ = 10 GeV/c; $\epsilon$ is the relative difference between FS and LUT.}
\newcolumntype{C}{>{\centering\arraybackslash}X}
\begin{tabularx}{\textwidth}{CCCccccccc}
\toprule
 \multirow[b]{1.78}{*}{\boldmath{$\eta$}} & \multicolumn{3}{c}{\boldmath{$CS_X$}} & \multicolumn{3}{c}{\boldmath{$CS_Y$}} & \multicolumn{3}{c}{\boldmath{$C_Q$\textbf{ [ke]}}} \\
 \cmidrule{2-10}
      & \textbf{FS}& \textbf{LUT} & \textbf{$\epsilon$ [\%]}& \textbf{FS} & \textbf{LUT} & \textbf{\textbf{$\epsilon$} [\%]}&\textbf{FS}& \textbf{LUT} & \textbf{$\epsilon$ [\%]} \\
      \midrule
      0 &1.097 &	1.055	&3.8 & 1.082	& 1.04 &	3.9		& 7.229 &	7.006 &	3.1 \\
      \midrule
     1  &1.217	& 1.422	& 16.8	&3.023	&2.988 &	1.2 & 10.37 &	10.95	&5.6\\
     \midrule
     1.4 & 2.07 &	2.046 &	1.2 & 4.015 &	3.989 &	0.65 & 
     14.29 &	15.05 &	5.3 \\
     \bottomrule
\end{tabularx}
\end{table}


\begin{table}[H]
\caption{\label{tab:closure_100GeV}Average cluster size in transverse 
($CS_X$) and bending ($CS_Y$) direction, and average cluster charge $C_Q$ for FS- and LUT-based events for different pseudorapidity values $\eta$ and $p_T$ = 100 GeV/c; $\epsilon$ is the relative difference between FS and LUT.}
\newcolumntype{C}{>{\centering\arraybackslash}X}
\begin{tabularx}{\textwidth}{CCCccccccc}
\toprule
 \multirow[b]{1.78}{*}{\boldmath{$\eta$}} & \multicolumn{3}{c}{\boldmath{$CS_X$}} & \multicolumn{3}{c}{\boldmath{$CS_Y$}} & \multicolumn{3}{c}{\boldmath{$C_Q$\textbf{ [ke]}}} \\
 \cmidrule{2-10}
      & \textbf{FS}& \textbf{LUT} & \textbf{$\epsilon$ [\%]}& \textbf{FS} & \textbf{LUT} & \textbf{\textbf{$\epsilon$} [\%]}&\textbf{FS}& \textbf{LUT} & \textbf{$\epsilon$ [\%]} \\
      \midrule
      0 & 1.079 &	1.042 &	3.4 & 1.075 &	1.033	 &3.9   & 7.334 &	7.252 &	1.1   \\
      \midrule
     1  & 1.243 & 1.424 &  14.5 	&3.011 & 3.013 & 0.07  & 10.39 &	10.68 &	2.8 \\
     \midrule
     1.4 & 2.068	& 2.053 & 0.72  & 4.022 &3.986 & 0.89  & 
     14.29 &	14.76&	3.3  \\
     \bottomrule
\end{tabularx}
\end{table}

The discrepancy reported in Section \ref{sec:closure_cluster_size} is evident 
for $\eta$ = 1 for all $p_T$ values in transverse cluster size $CS_X$.

In order to investigate these discrepancies between FS- and LUT-based  events for $\eta$ = 1 in transverse cluster size, more pseudorapidity values were explored for pions with $p_T$ = 100 GeV/c. 
Simulations were run at $\eta$ = 0.4, 0.8 and 1.2. 
The results are reported in Table \ref{tab:eta_variations}.

\begin{table}[H]
\caption{\label{tab:eta_variations}Average cluster size in transverse 
 direction $CS_X$, for different pseudorapidity values $\eta$ and $p_T$ = 100 GeV/c; $\epsilon$ is the relative difference between FS and LUT.}
\newcolumntype{C}{>{\centering\arraybackslash}X}
\begin{tabularx}{\textwidth}{CCCC}
\toprule
  \multicolumn{4}{c}{\boldmath{$CS_X$}} \\
   \midrule
  
     \boldmath{$\eta$} & \textbf{FS}& \textbf{LUT} &  \textbf{$\epsilon$ [\%]}  \\
      \midrule
      0 & 1.079 &	1.042 &	3.4    \\
      \midrule
      0.4 & 1.09 &	1.047 &	3.9 \\
      \midrule
      0.8 & 1.117	& 1.073 &	3.9  \\
      \midrule 
     1  & 1.243 & 1.424 &  14.5 	\\
     \midrule
     1.2 & 2.042 &	2.018 &	1.2  \\
     \midrule
     1.4 & 2.068	& 2.053 & 0.72   \\
     \bottomrule
\end{tabularx}
\end{table}

For these  values of $\eta$, the difference in the average cluster size in the transverse  direction between FS- and LUT-based events is again within 1-4\%. 

These results indicate that the discrepancy observed at $\eta$ = 1 is 
limited to a very narrow range around that value. In order to confirm  that this the divergence of the beam for simulations at $\eta$ = 1, it was increased to $\pm$10 mrad, 
corresponding to a pseudorapidity range: $\eta \in [0.99,1.01]$. The results 
are reported in Table \ref{tab:BeamDiv_variations}.

\begin{table}[H]
\caption{\label{tab:BeamDiv_variations}Average cluster size in the transverse direction ($CS_X$) for various beam divergences at $\eta = 1$ and $p_T$ = 100 GeV/c; $\epsilon$ denotes the relative difference between FS and LUT.}
\newcolumntype{C}{>{\centering\arraybackslash}X}
\begin{tabularx}{\textwidth}{CCcc}
\toprule
\multicolumn{1}{c}{} & \multicolumn{3}{c}{\boldmath{$CS_X$}} \\
\midrule
\textbf{Beam Divergence} (\boldmath{$x,y$}) (\textbf{Mrad}) & \textbf{FS} & \textbf{LUT} & \boldmath{$\epsilon$ [\%]} \\
\midrule

0, 0   & 1.224 & 1.428 & 17 \\
10, 0  & 1.344 & 1.449 & 8 \\
0, 10  & 1.547 & 1.551 & 0.2 \\
10, 10 & 1.548 & 1.554 & 0.4 \\
\bottomrule
\end{tabularx}
\end{table}

When the beam divergence is  increased in the $y$ direction, the level of agreement between LUT and FS events is much 
better than 1\%. 

In general, simulations with such beam divergence show an improvement in the agreement for all 
observables, reaching about 2 \% in $CS_Y$ and less than 0.1 \% in $C_Q$.   
Since charged particles from LHC collisions are distributed  uniformly in  $\eta$ from $-$2 to +2, the observed local discrepancy   poses no problem at all.


\section{Conclusions and Outlook}
\label{sec:conclusions}
The extreme data rates and fluences expected at the HL-LHC impose stringent constraints on both detectors and computing resources. Radiation damage effects will significantly impair the performance of silicon pixel detectors in the ATLAS Inner Tracker. Thus, it is crucial to accurately replicate such degradation in MC events, as demonstrated by the current pixel detector and the existing radiation damage digitizer.

The pursuit of a faster digitizer while maintaining precision led to the development of the LUT digitizer.
In a closure test, the new radiation damage digitizer demonstrated excellent agreement with fully simulated data across all ranges of pseudorapidity and transverse momentum. The discrepancy at $\eta=1$ is confined to a range of less than 1\% in pseudorapidity, as detailed in Section \ref{sec:closure_summary}.

In the near future, data from ITk pixel modules equipped with the new readout chip will become available, allowing for the comparison of predictions from LUT-based simulations with testbeam data.

The LUT method will also be extended to 3D sensors \cite{EPS2023Bomben}. For these sensors, the LUT of CCE as a function of the radial distance of the carriers from the central electrode should suffice, as the Lorentz deflection is negligible. Efforts are underway to model edge areas for all types of sensors (planar, 3D, and strips).

Preliminary tests indicate that the LUT radiation damage digitizer operates as fast as the digitizer without radiation damage corrections. This was somewhat expected, as the computational load is similar for both. As shown in Equation (\ref{eq:rprop}), for unirradiated sensors, $k=1$, $\Delta z=w-z$, and $\tan\theta_{LA}$ is constant, resulting in no algorithmic difference. This also confirms that reading values from LUTs is not the dominant factor in the computing load.

In summary, the LUT method for simulating radiation damage effects  is as precise as full simulation methods and as fast as algorithms 
that do not model radiation damage. The LUT method is expected to  be used as the default for the generation of MC events for the HL-LHC 
phase of ATLAS.



\vspace{6pt}
\authorcontributions{Keerthi Nakkalil implented the LUT algorithm and ran the Allpix$^2$ simulations. Keerthi Nakkalil wrote Sections 3 and most of 4. Marco Bomben oversaw the project, ran the TCAD simulations and wrote all the rest of the article.} 





\acknowledgments{The authors want to warmly thank Allpix$^2$ developers Simon  Spannagel and Paul Sch\"utze for their help in implementing 
the LUT algorithm and for many useful discussions.  The authors want also to thank Tomas Dado for the tests on computing 
performance of the algorithm. }




\abbreviations{Abbreviations}{
The following abbreviations are used in this manuscript:\\

\noindent 
\begin{tabular}{@{}ll}
LUT & Look-Up Tables\\
LHC & Large Hadron Collider\\
HL-LHC & high-luminosity LHC\\
ITk & Inner Tracker \\
CS & cluster size
\end{tabular}
}



\newpage
\begin{adjustwidth}{-\extralength}{0cm}

\reftitle{References}

\PublishersNote{}
\end{adjustwidth}
\end{document}